\documentclass[twocolumn,tighten]{aastex631}

\graphicspath{{./}{}}
\DeclareUnicodeCharacter{2212}{-}
\usepackage{multirow}
\usepackage{csvsimple,longtable,booktabs}
\usepackage{amsmath}
\usepackage{tikz}

\begin{document}

\submitjournal{ApJ}
\title{Critical Condition of Core-Collapse Supernovae I: One Dimensional Models}

\shorttitle{Critical Condition of Core-Collapse Supernovae}
\shortauthors{Pochik \& Thompson}
\correspondingauthor{David Pochik}
\email{pochik.1@buckeyemail.osu.edu}

\author[0000-0002-8310-0271]{David Pochik}
\affiliation{Center for Cosmology and Astroparticle Physics, 191 West Woodruff Avenue, Columbus, OH 43210, USA}
\affiliation{Department of Physics, Ohio State University, 191 West Woodruff Avenue
Columbus, OH 43210, USA}

\author[0000-0003-2377-9574]{Todd A. Thompson}
\affiliation{Center for Cosmology and Astroparticle Physics, 191 West Woodruff Avenue, Columbus, OH 43210, USA}
\affiliation{Department of Physics, Ohio State University, 191 West Woodruff Avenue Columbus, OH 43210, USA}
\affiliation{Department of Astronomy, Ohio State University, 180 West 18th Avenue, Columbus, OH 43210, USA}

\begin{abstract}
When the core of a massive star collapses, neutrino heating can energize the stalled accretion shock, leading to a successful supernova. The critical condition that characterizes the transition from accretion to explosion is a central topic of study and is often characterized by a critical proto-neutron star (PNS) neutrino luminosity $L_\nu^{\rm crit}$, which depends on the post-collapse mass accretion rate $\dot{M}$ from the progenitor. We examine the critical condition by solving the spherically symmetric time-dependent Euler equations with a general equation of state and realistic microphysics for a range of $\dot{M}$, average neutrino energy $\left< \epsilon_{\nu}\right>$, luminosity $L_{\nu}$, PNS radius $R_{\star}$, mass $M_{\star}$, and pre-shock Mach number $\mathcal{M}$ for a fixed neutrino optical depth from the PNS surface of $2/3$. We derive $L_{\nu}^{\mathrm{crit}}$ as a function of the input parameters. We show that pressurized pre-shock flow, as parameterized by low $\mathcal{M}$, changes the normalization of the critical condition because accretion of higher entropy shells at later times after collapse leads to lower  $L_{\nu}^{\mathrm{crit}}$. We connect this finding to the onset of explosion due to compositional interface accretion. Across our parameter space, we test critical conditions that have been proposed in the literature, including the ``antesonic" condition, the ``force explosion condition," and the heuristic heating-advection timescale condition. We discuss how shock oscillations impact these critical conditions. Compared to other explosion conditions, we find that the antesonic ratio shows the least variation across the model space we explore. This work is preparatory for similar experiments in 2D axisymmetry and 3D.
\end{abstract}



\section{Introduction}
\label{sec:introduction}
At the end of its life, a massive ($M\gtrsim 8 \, \mathrm{M}_{\odot}$) star's iron core rapidly collapses to high densities ($\gtrsim10^{14} \, \mathrm{g} \, \mathrm{cm}^{-3}$). The strong nuclear force becomes repulsive among the tightly packed nucleons, causing the dense core -- i.e., the protoneutron star (PNS) -- to ``bounce," and send a shockwave through the infalling material of the star \citep{Colgate1961}. The shock is stalled by energy losses via nuclear dissociation, neutrino cooling, and ram pressure \citep{Mazurek1982,Burrows1985,Bethe1990,Ott2018b}. Some physical mechanism re-energizes the shock and produces a core-collapse supernova (CCSN) \citep[see, e.g.,][for reviews]{Janka2012,Burrows2013,Burrows2021} accompanied by a neutrino-driven wind that emanates from the PNS surface \citep[see, e.g.,][]{Qian1996,Thompson2001,Prasanna2022}. If the mechanism fails, stellar material continues to accrete onto the PNS until a black hole (BH) forms (potentially observed by \cite{gerke2015,adams2017a,adams2017b,neustadt2021}; see also \cite{kochanek2008})

Understanding the CCSN mechanism and how it is related to neutron star (NS) and progenitor properties is a central focus in astrophysics. A conventional explanation is the so-called ``delayed neutrino mechanism" \citep{Bethe1985,Bruenn1995}, which continues to be studied with 3D simulations \citep{Takiwaki2014,Lentz2015,Melson2015,Janka2016,Muller2017,Ott2018b,Summa2018,Burrows2020a,Stockinger2020, Burrows2021, WangBurrows2024}. Because these types of simulations are computationally expensive \citep{couch2017,Mezz2020}, studies are sometimes done with parameterized models with simplified neutrino schemes in 1D and 2D \citep{Burrows1993,Scheck2006,Kotake2007,Murphy2008,Nordhaus2010,Hanke2012,Couch2015}, which still capture important physical properties while being much less computationally intensive. 

Through the development of these simplified models, various critical conditions for explosion have been explored, which are used to distinguish exploding from non-exploding CCSN models \citep{Burrows1993,Pejcha2012,Murphy2008,Fernandez2012,Raives2018,Murphy2017,Raives2021,Gogilashvili2022, Gogilashvili2023}.
Specifically, \cite{Burrows1993}, demonstrated the existence of a critical condition for the neutrino  mechanism using a spherically symmetric, time-steady configuration for the accretion flow. They showed that there exists a critical neutrino luminosity $L_{\nu}^{\mathrm{crit}}$ above which no stable accretion solution exists. Since then, many studies have been performed on $L_{\nu}^{\mathrm{crit}}$ in 1D, 2D, and 3D. These studies have revealed a nontrivial dependence of $L_{\nu}^{\mathrm{crit}}$ on PNS rotation and the EoS \citep{Yamasaki2005,Yamasaki2006}, 1D shock dynamics \citep{Fernandez2012,Gabey2015}, nuclear dissociation energy \citep{Fernandez2014,2015Fernandez}, and spatial dimensionality with hydrodynamic instabilities \citep{Murphy2008, Nordhaus2010, Hanke2012, Couch2013, Mabanta2018}. 

Several explanations for the physics of the critical luminosity also been proposed. One example is the heuristic timescale criterion \citep[e.g.,][]{Thompson2005}, which states that if the time required for matter to advect through the gain region is longer than the timescale to heat the matter in the gain region, then an explosion occurs. Another is the ``antesonic condition" \citep{Pejcha2012, Raives2018, Raives2021}, which states that for free-fall onto a standing shockwave no time-steady accretion solutions exist if the square of the ratio of the post-shock sound speed to the escape velocity exceeds a critical value. \cite{Pejcha2012} show that this physics explains the existence of $L_{\nu}^{\mathrm{crit}}$ from \cite{Burrows1993}. Another explanation for the physics of $L_{\nu}^{\mathrm{crit}}$ is the force explosion condition (FEC) \citep{Murphy2017,Gogilashvili2022,Gogilashvili2023}, which evaluates the net integral of forces acting on the shock to determine the analytic criterion for explosion.

We investigate the physics of the 1D critical condition by developing parameterized CCSN accretion models in 1D with boundary conditions tuned to model pressurized inflow. We use model inputs that capture key details of the accretion phase and allow us to study the physics of criticality. We prescribe the PNS neutrino luminosity $L_{\nu}$, neutrino spectra $\left< \epsilon_{\nu} \right>$, mass accretion rate $\dot{M}$, PNS radius $R_{\star}$, PNS mass $M_{\star}$, neutrino optical depth $\tau$, and pre-shock Mach number $\mathcal{M}$ as inputs for our models. The latter captures the thermal content of the pre-shock, infalling stellar material. Note that 1D accretion models are typically constructed with pressureless or near-pressureless free-fall ($\mathcal{M} \gg 1$) in the pre-shock region \citep{Pejcha2012,Fernandez2012,Gabey2015,Raives2018,Raives2021}. However, the fluid above the shock has non-zero pressure in realistic progenitors \citep{Woosley2002,Woosley2007}. As an example, \cite{Pejcha2012} showed that the antesonic condition is reduced when the pre-shock fluid is pressurized, thus the pre-shock pressure may directly affect condition(s) for explodability.  

In particular, previous works show that when the Si/O layer accretes onto the stalled shock, the ram pressure $P_{\mathrm{ram}}$ and $\dot{M}$ decrease, which either drives the model closer to criticality or directly leads to explosion \citep{Ott2013,Ott2018a,Ott2018b, vartanyan2018, Wang2022}. However, changes in the thermal content of the accreting material may effect criticality as well. Accretion of mass with higher pressure will lead to smaller infall velocity at the shock. Both will affect explodability \citep{Pejcha2012}. The changes in $P_{\mathrm{ram}}$ have been highlighted in Si/O layer studies, but the changes in the thermal content, such as the entropy $S$ of the infalling material, have received less attention. These changes may map into the critical condition in a non-trivial way. It is possible that both the change in $P_{\mathrm{ram}}$ and $S$ are connected to the transition from accretion to explosion when the Si/O interface accretes onto the shock. 

Here, we extend the analysis performed by previous studies. Our primary goals are (1) to understand the physics of various explosion conditions and their connections to the neutrino heating mechanism,
(2) to understand why the critical condition changes as a function of the controlling parameters, and (3) to prepare a framework for future 2D and 3D critical condition studies.  Prior studies show that the normalization of the critical curve depends on $\tau$ \citep{Fernandez2012}, neutrino properties, and PNS properties \citep{Pejcha2012}, but they do not show why the critical condition depends on these and other parameters. Moreover, some accretion studies \citep{Pejcha2012,Gabey2015,Raives2018,Raives2021} have not captured the effects of $\mathcal{M}$ on the shock dynamics. We investigate the role of both the finite pre-shock pressure, i.e., low $\mathcal{M}$, and other controlling parameters on criticality in our time-dependent models.

In Section \ref{sec:code}, we introduce the numerical tools for this study. Section \ref{subsec:hydro} introduces the hydrodynamics equations. The governing equations for neutrino-matter interactions and composition rates are covered in Section \ref{subsec:micro}. The boundary and initial conditions are covered in Sections \ref{subsec:BC} and \ref{subsec:IC}, respectively. Section \ref{sec:results1D} begins our discussion on results. Specifically, Section \ref{subsec:Fiducial} shows our fiducial accretion models with their dependence on $\mathcal{M}$ and  $L_{\nu}$. Section \ref{subsec:CritCurve} introduces our procedure for finding the critical curve and includes a discussion of shock dynamics, and Section \ref{subsubsec:CritCurveDepOnInputs} covers the functional dependence of $L_{\nu}^{\mathrm{crit}}$ on the controlling parameters. Section \ref{subsec:Antesonic} covers our results on the antesonic condition, Section \ref{subsec:advheat} shows our results on the advection and heating timescales, and Section \ref{subsec:FEC} shows our findings on the force explosion condition (FEC). Section \ref{subsec:MachNumberRealisticProgenitors} discusses the potential importance of $\mathcal{M}$ in criticality studies that use realistic progenitors. We conclude with Section \ref{sec:conclusions}.

\section{Numerical Framework}
\label{sec:code}
We use the 3D radiative magnetohydrodynamics (MHD) code Athena++ \citep{Stone2020}, including a general equation of state \citep{Timmes2000,Coleman2020} and neutrino microphysics \citep[see, e.g.,][]{Scheck2006}. All calculations here are spherically symmetric, and spatial discretization is performed in Athena++ with a second-order piecewise parabolic method \citep{Colella1984}. The fluxes at cell boundaries are determined with the Harten-Lax-van Leer-Contact (HLLC) \citep{Toro1994} Riemann solver. For temporal integration, we use the second-order accurate van Leer predictor-corrector algorithm \citep{Stone2009} with the corner transport upwind method of \cite{Collela1990} along with a second-order strong-stability preserving Runge-Kutta method constrained by a Courant-Friedrichs-Lewy number \citep[see, e.g.,][]{DeMoura2013} of $\mathrm{CFL}=0.3$. We discretize the computational domain into $n_{R}=256$ radial zones with logarithmic spacing. We observe changes in $\displaystyle{L_{\nu}^{\mathrm{crit}}}$ on the order of $1\%$ or less when changing $n_{R}$ between values of $128$ to $512$, thus we do not examine other radial resolutions for the studies outlined in this paper.

\subsection{Hydrodynamics}
\label{subsec:hydro}
We solve the fully compressible Euler equations for the conservation of mass, momentum, and energy:
\begin{equation}
\label{eqn:MassConservation}
    \frac{\partial \rho}{\partial t} + {\nabla} \cdot \left(\rho \, \bf{v}\right)=0,
\end{equation}
\begin{equation}
\label{eqn:MomentumConservation}
    \frac{\partial \left(\rho \bf{v}\right)}{\partial t} + \mathbf{\nabla} \cdot \left(\rho \mathbf{v} \otimes \mathbf{v}\right) + \nabla P = -\rho \frac{G M_{\star} }{r^2} \hat{r},
\end{equation}
\begin{equation}
\label{eqn:EnergyConservation}
    \frac{\partial E}{\partial t} + \nabla \cdot \left[ \left(E+P\right)\bf{v}\right] = \mathcal{L}_{\mathrm{net}},
\end{equation}
where $\rho$ is the mass density, $\mathbf{v}$ is the velocity vector, $P$ is the pressure, $r$ is the radius, $G$ is Newton's constant, $E=\rho \epsilon$ is the total energy density, and $\mathcal{L}_{\mathrm{net}}=\rho\dot{Q}=\rho\left(\dot{Q}_{\mathrm{H}}-\dot{Q}_{\mathrm{C}}\right)$ is the net energy deposition rate per unit volume due to neutrino-matter interactions.

\subsection{Microphysics}
\label{subsec:micro}
The PNS is excised and replaced with an inner boundary at $R_{\star} \geq 30 \, \mathrm{km}$ in our calculations, therefore $\rho\lesssim 10^{12} \, \mathrm{g} \, \mathrm{cm}^{-3}$ near the inside boundary (see Section \ref{subsec:BC} for details on inner boundary conditions). We do not include an EoS for dense nuclear matter because of the relatively low inner boundary $\rho$, but we still require an EoS that captures the transition for electrons ($e^{-}$) and positrons ($e^{+}$) from relativistic to non-relativistic regimes since the temperature $T$ varies from a few MeV to tenths of an MeV for $r \in \left[R_{\star}, \, 1000 \, \mathrm{km}\right]$. Using the general EoS framework in Athena++ developed by \cite{Coleman2020}, equations \ref{eqn:MassConservation}, \ref{eqn:MomentumConservation}, and \ref{eqn:EnergyConservation} are closed with a general EoS for $e^{-}$, $e^{+}$, photons, and non-relativistic baryons based on tabular interpolation of the Helmholtz free energy \citep{Timmes2000}.

The electron-type charged current processes dominate energy exchange and electron fraction $Y_{e}$ evolution in the accretion flow \citep{Janka2000}:
\begin{equation}
\label{eqn:ChargedCurrentElectron}
    e^{-}+p \leftrightarrow n+\nu_{e},
\end{equation}
\begin{equation}
\label{eqn:ChargedCurrentPositron}
    e^{+}+n \leftrightarrow p+\bar{\nu}_{e}.
\end{equation}
We spatially and temporally resolve $Y_{e}$ by solving the source function for electron-type charged current processes \citep{Qian1996}:
\begin{equation}
\label{eqn:SourceFunction}
\begin{split}
    v_{r} \frac{dY_{e}}{dr} = \lambda_{\nu_{e}n}+\lambda_{e^{+}n}- \\ \left(\lambda_{\nu_{e}n}+\lambda_{e^{+}n}+\lambda_{\bar{\nu}_{e}p}+\lambda_{e^{-}p}\right)Y_{e},
\end{split}
\end{equation}
where $v_{r}=\left|\mathbf{v}\right|$ in spherical symmetry. The forward and reverse reaction rates are provided by \cite{Qian1996} (in units of $\mathrm{s}^{-1}$)\footnote{See Appendix D in \cite{Scheck2006} for a detailed derivation for both the reaction rates and neutrino heating/cooling rates covered in this section.}:
\begin{equation}
\label{eqn:foward1}
\begin{split}
    \lambda_{\nu_{\rm e} n}\approx \frac{1+3\alpha^2}{2\pi^2}\left(\hbar c\right)^2G_{\rm F}^2 \frac{L_{\rm \nu_e}}{R_{\star}^2} \times \\ \left(\epsilon_{\nu_{e}} +2\Delta + \frac{\Delta^2}{\left<\epsilon_{\nu_{e}}\right>}\right)(1-x),
\end{split}
\end{equation}
\begin{equation}
\label{eqn:forward2}
\begin{split}
    \lambda_{\bar{\nu}_{\rm e} p}\approx \frac{1+3\alpha^2}{2\pi^2}\left(\hbar c\right)^2G_{\rm F}^2 \frac{L_{\rm \bar{\nu}_e}}{R_{\star}^2} \times \\ \left(\epsilon_{\bar{\nu}_{e}} -2\Delta + \frac{\Delta^2}{\left<\epsilon_{\bar{\nu}_{e}}\right>}\right)(1-x),
\end{split}
\end{equation}
\begin{equation}
\label{eqn:reverse1}
    \lambda_{e^{-}p} \approx 0.448 \ T_{\rm MeV}^5 \  \frac{F_4(\eta_{\rm e})}{F_4(0)},
\end{equation}
\begin{equation}
\label{eqn:reverse2}
    \lambda_{e^{+}n} \approx 0.448 \ T_{\rm MeV}^5 \  \frac{F_4(-\eta_{\rm e})}{F_4(0)},
\end{equation}
where $\alpha=1.26$ is the axial coupling vector, $G_{\mathrm{F}}^2=1.16637\times 10^{-11} \, \mathrm{MeV}^{-2}$ is the Fermi constant, $\hbar$ is the reduced Planck's constant, $c$ is the speed of light, $\epsilon_\nu=\left<\epsilon_{\nu}^{2}\right>/\left<\epsilon_{\nu}\right>$  is a ratio of neutrino energy moments, $L_{\nu}$ is the neutrino luminosity, $\Delta=1.2935 \, \mathrm{MeV}$ is the neutron-proton mass difference, $\eta=\mu/k_{B}T$ is the degeneracy parameter of a given species with chemical potential $\mu$, $x=\sqrt{1-{R_{\star}^2}/{r}^2}$, $T_{\mathrm{MeV}}=T/1 \,\mathrm{MeV}$, and $F_{n}\left(\eta\right)$ is the Fermi integral,
\begin{equation}
\label{eq:FermiIntegral}
    F_{n}\left(\eta\right)=\int_{0}^{\infty} \, dx \, \frac{x^n}{\mathrm{exp}\left[x-\eta\right]+1}.
\end{equation}
The neutrino energy moments in general are given by \citep{Thompson2001}:
\begin{equation}
\label{eqn:NeutrinoEnergyMoment}
    \left<\epsilon_{\nu}^n\right>=\frac{\int_{-1}^{1}d \mu \int \, \epsilon_{\nu}^{n} f_{\nu}\left(r,t,\epsilon_{\nu}, \mu\right) d^3\epsilon_{\nu}}{\int_{-1}^{1}d \mu \int \, f_{\nu}\left(r,t,\epsilon_{\nu_{\eta}}, \mu\right)d^3\epsilon_{\nu}},
\end{equation}
where,
\begin{equation}
\label{eqn:FermiTerm}
\begin{split}
    f_{\nu}\left(r,t,\epsilon_{\nu},\mu\right)=g_{\nu}\left(r,t,\mu\right) \times \\ \left[1+\mathrm{exp}\left(\epsilon_{\nu}/k_{b}T_{\nu}-\eta_{\nu}\right)\right]^{-1},
\end{split}
\end{equation}
and $g_{\nu}$ is an angle-dependent function \citep{Scheck2006}. Equation \ref{eqn:NeutrinoEnergyMoment} can be simplified to show how $\left< \epsilon_{\nu_{e}} \right>$ depends on the neutrino temperature $T_{\nu}$ \citep{Scheck2006}:
\begin{equation}
\label{eqn:NeutrinoEnergyMomentSimplified}
    \left<\epsilon_{\nu}^{n}\right> = \left(k_{B}T_{\nu}\right)^{n} \frac{F_{n+2}\left(\eta_{\nu}\right)}{F_{2}\left(\eta_{\nu}\right)},
\end{equation}
where we set $\eta_{\nu_{e}}=\eta_{\bar{\nu}_{e}}=0$. The net neutrino heating rate per unit mass $\dot{Q}=\dot{Q}_{\mathrm{H}}-\dot{Q}_{\mathrm{C}}$ is given by modified versions of the \cite{Qian1996} rates (in units of $\mathrm{MeV} \, \mathrm{s}^{-1} \mathrm{g}^{-1}$),
\begin{equation}
\label{eqn:heatingrate}
\begin{split}
    \dot{Q}_{\mathrm{H}} \approx 9.65 \, N_{A} \big(X_{n} \, L_{\nu_{e},51}\, \epsilon_{\nu_{e},\mathrm{MeV}}^{2} \\ + X_{p} \, L_{\bar{\nu}_{e},51}\, \epsilon_{\bar{\nu}_{e},\mathrm{MeV}}^{2}\big)\frac{1-x}{R_{\star,6}^2},
\end{split}
\end{equation}
\begin{equation}
\label{eqn:coolingrate}
    \dot{Q}_{\mathrm{C}} \approx \frac{2.27 \, N_{A}}{F_{5}\left(0\right)}T_{\mathrm{MeV}}^6\left(X_{p} \, F_{5}\left(\eta_{e}\right)+ X_{n} \, F_{5}\left(-\eta_{e}\right)\right),
\end{equation}
where $N_{A}$ is Avogadro's constant,  $X_{p}=Y_{e}$ is the proton fraction, $X_{n}=1-Y_{e}$ is the neutron fraction, $L_{\nu,51}=L_{\nu}/10^{51} \, \mathrm{ergs} \, \mathrm{s}^{-1}$, $\epsilon_{\nu,\mathrm{MeV}}^2={\left<\epsilon_{\nu}^3\right>} /{\left<\epsilon_{\nu}\right>}/ 1 \, \mathrm{MeV}$, and $R_{\star,6}=R_{\star}/10^{6} \, \mathrm{cm}$. We set $L_{\nu_{e}}=L_{\bar{\nu}_{e}}=L_{\nu}$ everywhere, and we modify the rates of \cite{Qian1996} by implementing $\eta_{e} \neq 0$ where applicable, e.g., Equations \ref{eqn:reverse1}, \ref{eqn:reverse2}, and \ref{eqn:coolingrate} \citep{Thompson2001}. Note that the post-shock radius at which $\dot{Q}>0$ is the gain radius $R_{\mathrm{gain}}$, and the area between $R_{\mathrm{gain}}$ and the shock radius $R_{\mathrm{shock}}$ is the ``gain region" \citep{Bethe1985}. Above $R_{\mathrm{shock}}$, the charged-current interactions (equations \ref{eqn:ChargedCurrentElectron} and \ref{eqn:ChargedCurrentPositron}) are suppressed because of the low free-nucleon fraction of the infalling material. Primarily $Y_{e}=26/56 \sim 0.46$ material (Fe$^{56}$ nuclei) fall onto the shock during accretion, which changes to $Y_{e} \sim 0.5$ once the Si/O layer accretes onto the shock (see Section \ref{subsec:MachNumberRealisticProgenitors} for more details on the Si/O layer). The compositional changes across the shock are not inherently captured in the Helmholtz EoS, thus we modify the heating/cooling and reaction rates with the free nucleon mass fraction from \cite{Qian1996} to model the suppression in $\dot{Q}$ at  $r>R_{\mathrm{shock}}$, \begin{equation}
\label{eqn:NucleonMassFraction}
    \chi_{N} = 828 \, \frac{T_{\mathrm{MeV}}^{9/8}}{\rho_{8}^{3/4}} \, \mathrm{exp}\left(-\frac{7.074}{T_{\mathrm{MeV}}}\right),
\end{equation}
where $\rho_{8}=\rho/10^{8} \, \mathrm{g} \, \mathrm{cm}^{-3}$. Equations \ref{eqn:heatingrate} and \ref{eqn:coolingrate} are multiplied by $\mathrm{min}\left(\chi_{N}, \, 1\right)$. The steep drop in $T$ at $r>R_{\mathrm{shock}}$ causes $\chi_{N}$ to rapidly decrease, which approximately models the drop in $\dot{Q}$ across the shock. This treatment is effective for a wide parameter space, but it sometimes does not effectively decrease the pre-shock $\dot{Q}$. In some regimes, the pre-shock $\dot{Q}$ slowly drops to zero instead of being immediately driven to zero by $\chi_{N}$,  as shown in Figures \ref{fig:VaryMachNumberMultiPanel}, \ref{fig:PT2012_ReplicatePlot}, and \ref{fig:EntropyProfilesProgenitors}. Finally, following 
\citep{Fernandez2012}, we suppress the heating, cooling, and $Y_{e}$ number rates at high density using:
\begin{equation}
\label{eqn:exponentialsuppression}
    f_{\mathrm{sup}}=\mathrm{exp}\left(-\rho / \rho_0 \right),
\end{equation}
where $\rho_0$ is the inner boundary density from the initial condition (IC) profile.

\subsection{Boundary Conditions}
\label{subsec:BC}
Since we implement a second-order spatial integration method in Athena++, we apply two ghost zones at the inner (denoted as 'gi') and outer (denoted as 'go') boundaries of the computational domain. For simplicity, $\rho$, $Y_{e}$, $v_{r}$, $T$, and $P$ are constant across the ghost zones for a given time step. We enforce inflow boundary conditions to model accretion, e.g., $Y_{e,\mathrm{gi}}$ and $v_{r,\mathrm{gi}}$ are equal to their respective first active zone quantities. We compute $T_{\mathrm{gi}}$ by finding an equilibrium temperature $T_{\mathrm{eq}}$ that satisfies $\dot{Q}\left(\rho,T_{\mathrm{eq}},Y_{e}\right) \sim 0$ in the first active zone, then we set $T_{\mathrm{gi}}=T_{\mathrm{eq}}$.

It is important that we maintain constant $\tau$ for all stable models, because the normalization of the critical curve is a function of $\tau$ \citep{Fernandez2012}. We define $\tau$ as:
\begin{equation}
\label{eqn:NeutrinoOpticalDepth}
    \tau = \int_{R_{\star}}^{R_{\mathrm{out}}} dr \, N_{A} \sigma_{\nu_{e}} X_{n} \rho,
\end{equation}
where $R_{\mathrm{out}}$ is the outer boundary, and,
\begin{equation}
\label{eqn:ElectronNeutrinoCrossSection}
    \sigma_{\nu_{e}}=\sigma_{0} \left(\frac{1+3\alpha^2}{4}\right)\frac{\left(\left<\epsilon_{\nu_{e}}^{2}\right>+2 \Delta \left<\epsilon_{\nu_{e}}\right>+\Delta^2\right)}{\left(m_{e} c^2\right)},
\end{equation}
is the electron neutrino cross section \citep{Scheck2006} with $\sigma_0 \sim 1.76 \times 10^{-44} \, \mathrm{cm}^2$. Because $\rho$ drops significantly across the shock, the flow at $r<R_{\mathrm{shock}}$ dominates the integral for $\tau$ and the upper limit reduces from $R_{\mathrm{out}}$ to $R_{\mathrm{shock}}$. Given our assumption of optically thin neutrino heating and cooling, the neutrino optical depth up to the shock is $\tau \sim 2/3$. We maintain this value by manipulating the ghost-zone density values. First, we set a fixed density value $\rho_0$ in the ghost zones and calculate $\tau$. Once the physical time sufficiently exceeds the post-shock sound crossing time, i.e., $t \gtrsim t_{\mathrm{sound}} \sim r/c_{s} \sim 10^{-2} \, \mathrm{s}$, $\tau$ is checked via Equation \ref{eqn:NeutrinoOpticalDepth}. If $\tau \leq 2/3 - \delta_{\tau}$ or $\tau \geq 2/3 + \delta_{\tau}$, where $\delta_{\tau} = 2.0 \times 10^{-3}$, $\rho_{\mathrm{gi}}$ is increased or decreased by $\left(1+3.0 \times 10^{-3}\right)$ times its current value. Once another sound crossing time passes and the post-shock flow has settled from the perturbation in $\rho_{\mathrm{gi}}$, equation \ref{eqn:NeutrinoOpticalDepth} is recalculated and $\tau$ is checked again. This process iterates until the threshold $2/3 - \delta_{\tau} < \tau < 2/3 + \delta_{\tau}$ is met. This results in a time-steady numerical simulation with $\tau \sim 2/3$ and allows us to consistently map the critical curve. 

At the outer boundary, $R_{\mathrm{out}}=1000 \, \mathrm{km}$, $\rho_{\mathrm{go}}$, $Y_{e,\mathrm{go}}$, and $v_{r,\mathrm{go}}$ are set equal to their respective values from the steady state IC profile, whereas $T_{\mathrm{go}}$ is set equal to the final active zone temperature. $P_{\mathrm{go}}$ is set with the Mach number $\mathcal{M}=v_{r}/c_{s}$ \citep{Fernandez2009}:
\begin{equation}
\label{eqn:PressureOuterboundary}
    P_{\mathrm{go}}=\frac{\rho_{\mathrm{go}}\left(\epsilon_{\mathrm{go}} + c^2\right)}{\frac{\Gamma_{\mathrm{go}} \mathcal{M}^2 c^2}{v_{r,\mathrm{go}}^2}-1} \sim \frac{\rho_{\mathrm{go}} v_{r,\mathrm{go}}^2}{\Gamma_{\mathrm{go}} \mathcal{M}^{2}},
\end{equation}
where $\Gamma_{\mathrm{go}}=\Gamma\left(\rho, \, T, \, Y_{e}\right)_{\mathrm{go}}$ is the adiabatic index, and $\epsilon_{\mathrm{go}}=\epsilon\left(\rho, \, T, \, Y_{e}\right)_{\mathrm{go}}$ is the total specific energy. Both quantities are interpolated from the EoS using the ghost zone quantities. $\mathcal{M}>1$ is a free parameter that controls the thermal content of the accreted matter, which directly affects the normalization of $L_{\nu}^{\mathrm{crit}}$ (Section \ref{subsubsec:CritCurveDepOnInputs}). As $\mathcal{M}$ increases above unity, the pre-shock flow becomes super-sonic and $P$ (equation \ref{eqn:PressureOuterboundary}) decreases. For $\mathcal{M} \gg 1.0$, $P$ ahead of the shock becomes negligible and the pre-shock flow approaches free-fall. Note that super-sonic flow must always be maintained at the outer boundary for accretion models, otherwise near- or sub-sonic flow becomes disrupted by pressure waves and prevents the fine-tuning of $\dot{M}$. 

We specify $\dot{M}$ with the cell-centered\footnote{Note that we are using cell-centered fluid variables here, but $R_{\mathrm{out}}=1000 \, \mathrm{km}$ is a face-valued quantity. We find this inconsistency to be quantitatively negligible in our critical condition results.} outer boundary density $\rho_{\mathrm{out}}$ and velocity $v_{r,\mathrm{out}}$ from the IC profile via $\dot{M}=4 \pi \rho_{\mathrm{out}} R_{\mathrm{out}}^2 \left|v_{r,\mathrm{out}}\right|$. We keep $L_{\nu}$ and $\left<\epsilon_{\nu}\right>$ constant as the models evolve, and these parameters are manipulated only at the start of simulations. Note that we have tested our models with a feature that slowly increases $L_{\nu}$ during the simulation and observed no qualitative change in $L_{\nu}^{\mathrm{crit}}$ compared to our current procedure.

\subsection{Initial Condition}
\label{subsec:IC}
We initialize our models with a numerical steady-state accretion profile generated with Athena++. Following previous works \citep[see, e.g.,][]{Fernandez2012,Pejcha2012}, we select values for $P$ ($\sim 10^{30} \, \mathrm{ergs} \, \mathrm{cm}^{-3}$), $\rho$ ($\sim 10^{11} \, \mathrm{g} \, \mathrm{cm}^{3}$), $Y_{e}$ ($\sim 1.0 \times 10^{-1}$), $T$ ($\sim 4 \, \mathrm{MeV}$), and $v_{r}$ ($-10^{8}$ cm s$^{-1}$) that are appropriate for the PNS surface. The outer boundary $P$ ($\sim 10^{24} $ ergs cm$^{-3}$), $\rho$ ($10^{7}$ g cm$^{-3}$), $Y_{e}$ ($\sim 26/56$), $T$ ($\sim 0.1 \, \mathrm{MeV}$), and $v_{r}$ ($-10^{9}$ cm s$^{-1}$) are selected to match the ``pre-shocked" quantities in realistic progenitors \citep{Woosley2002,Woosley2007}. The stalled shock is located at $50 \, \mathrm{km} \lesssim R_{\mathrm{shock}} \lesssim 200 \, \mathrm{km}$, which depends on the input physics. Our fiducial models use inputs of $R_{\star}=30 \, \mathrm{km}$, $M_{\star}=1.4 \, \mathrm{M}_{\odot}$, $\left< \epsilon_{\nu_{e}}  \right> \simeq 12.6 \, \mathrm{MeV}$, and $\left<\epsilon_{\bar{\nu}_{e}}\right> \simeq 18.9 \, \mathrm{MeV}$, which correspond to $T_{\nu_{e}}=4 \, \mathrm{MeV}$ and $T_{\bar{\nu}_{e}}= 6 \, \mathrm{MeV}$ respectively \citep{Fernandez2012}. We maintain $\left<\epsilon_{\bar{\nu}_{e}}\right> / \left<\epsilon_{\nu_{e}}\right> = 1.5$ throughout our analysis, and we keep the average energies constant when varying $L_{\nu}$. We vary the fiducial parameters to understand how the critical condition depends on PNS properties (see Sections \ref{subsec:CritCurve} and \ref{subsubsec:CritCurveDepOnInputs}) and $\mathcal{M}$ (see Sections \ref{subsec:Antesonic}, \ref{subsec:advheat}, and \ref{subsec:FEC}).

We generate a numerical accretion model (see Section \ref{subsec:Fiducial}) and evolve it in Athena++. An initial transient occurs and the model is driven toward time-steady flow. We evolve non-exploding models up to $t=3 \, \mathrm{s}$ to ensure stability. Steady-state accretion is defined by $\tau \sim 2/3$ (see Section \ref{subsec:BC}) with a stationary shock. Our definition of a ``stationary" shock is one that maintains either (1) $V_{\mathrm{shock}}\left(t\right)=0$ (stable, non-oscillatory) or (2) $V_{\mathrm{shock}}\left(t\right)\neq 0$ but $\left<V_{\mathrm{shock}}\right>_{t} \sim 0$ (stable, oscillatory, see Section \ref{subsec:CritCurve}). If $V_{\mathrm{shock}}$ becomes non-zero or changes sign anytime after $t=2.4 \, \mathrm{s}$ but does not explode, the model is flagged as stable and oscillatory. This is similar to the work done by \cite{Gabey2015}, where they observe oscillatory, but stable accretion models at low $\dot{M}$. Once a stable model fulfills these criteria, it is used as a ``checkpoint" in the $L_{\nu}-\dot{M}$ plane, and we progress to a new $L_{\nu}$ or $\dot{M}$ by using the stabilized model as the new IC profile. At constant $\dot{M}$, we increase $L_{\nu}$ step by step, which results in either a new steady state configuration or an explosion. The latter is characterized by the moment when the shock advects through the outer boundary and $v_{r,\mathrm{out}}>0$.

\begin{figure*}
\centering{}
\includegraphics[width=0.8\linewidth]{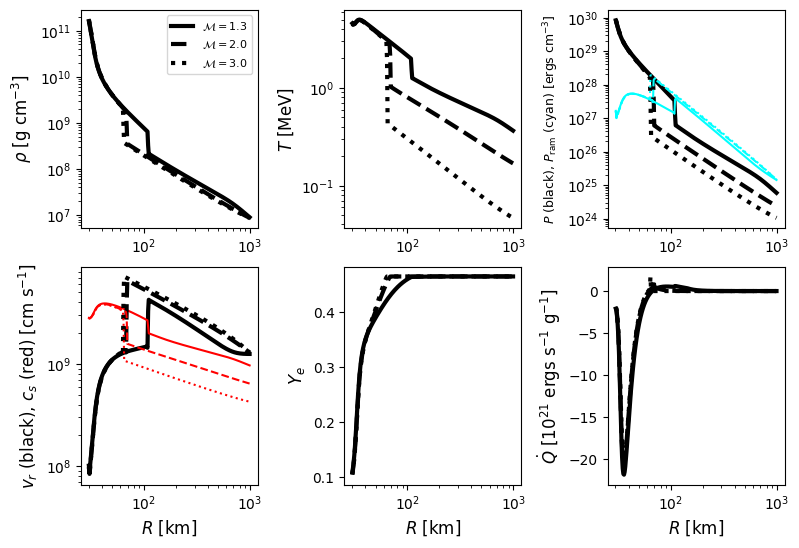}
\caption{Steady state profiles with fiducial inputs at $L_{\nu}=30 \times 10^{52}$ and $\dot{M}=0.7$ M$_{\odot}$ s$^{-1}$ for $\rho$, $T$, $P$, $v_{r}$, $Y_{e}$, and $\dot{Q}$ using $\mathcal{M}=1.3$ (solid), $2.0$ (dashed), and $3.0$ (dotted). The red profiles in the bottom left panel are $c_{s}$, and the cyan profiles in the top right panel are $P_{\mathrm{ram}}$. $R_{\mathrm{shock}}$ is characterized by the strong discontinuity in the profiles, which is where the accretion flow transitions from super- to sub-sonic. Due to the strong $\sim T^6$ dependence in $\dot{Q}_{C}$, the steep drop in $T$ across the shock leads to a sharp increase in $\dot{Q}$ just behind the shock. Because the EoS does not intrinsically capture the transition between nuclei and free nucleons across the shock, we use $\chi_{N}$ to reduce $\dot{Q}$ to zero at $r>R_{\mathrm{shock}}$. However, this treatment is less effective for $\mathcal{M}=1.3$, where $\dot{Q}$ slowly approaches zero in the pre-shock region}
\label{fig:VaryMachNumberMultiPanel}
\end{figure*}

\begin{figure*}
\centering{}
\includegraphics[width=0.70\linewidth]{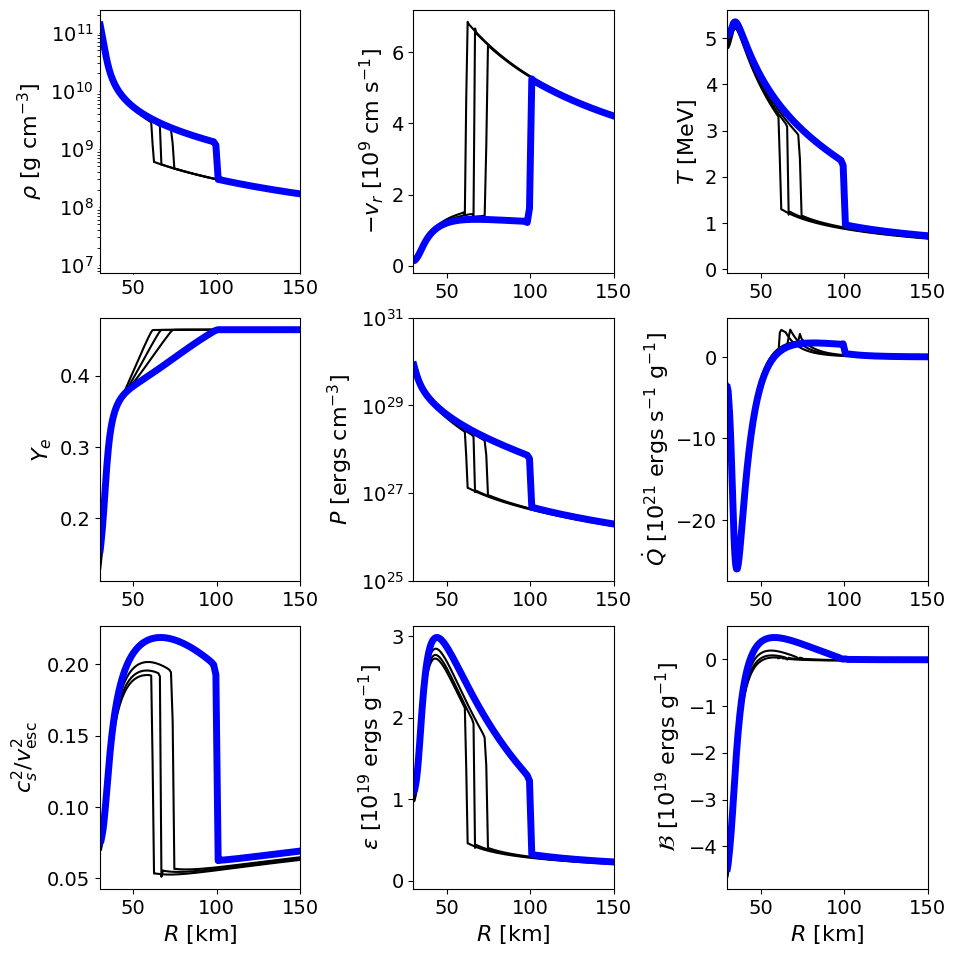}
\caption{Time steady solutions for $\rho$, $v_{r}$, $T$, $Y_{e}$, $P$, $\dot{Q}$, the antesonic ratio $c_{s}^2/v_{\mathrm{esc}}^2$, $\epsilon$, and the Bernoulli integral $\mathcal{B}$ from a series of simulations using $\dot{M}=1.0 \, \mathrm{M}_{\odot} \, \mathrm{s}^{-1}$ and $\mathcal{M}=2.0$ with fiducial inputs. The black profiles show model evolution as $L_{\nu}$ is increased over the range $\left[50, \, 70 \right] \times 10^{51} \, \mathrm{ergs} \, \mathrm{s}^{-1}$ in increments of $10^{51} \, \mathrm{ergs} \, \mathrm{s}^{-1}$. The blue profile is evaulated at $L_{\nu}=77 \times 10^{51} \, \mathrm{ergs} \, \mathrm{s}^{-1}$, which is just beneath $L_{\nu}^{\mathrm{crit}}$. The bottom right panel shows positive $\mathcal{B}$ values in the post-shock region for stable models. For the lower $L_{\nu}$ models shown here (black profiles), $\chi_{N}$ does not fully suppress $\dot{Q}$ in the immediate post-shock region, causing $\dot{Q}$ to smoothly approach $0$ at larger radii.} 
\label{fig:PT2012_ReplicatePlot}
\end{figure*}

\begin{figure}
\centering{}
\includegraphics[width=0.9\linewidth]{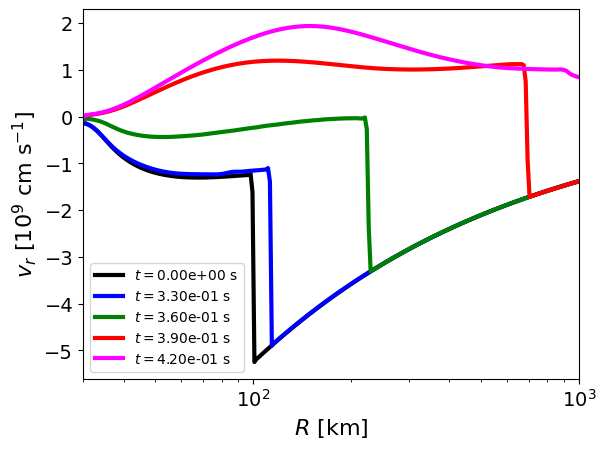}
\caption{$v_{r}$ profiles for the exploding model, $L_{\nu}^{\mathrm{crit}}=77.2 \times 10^{51} \, \mathrm{ergs} \, \mathrm{s}^{-1}$, from the series shown in Figure \ref{fig:PT2012_ReplicatePlot}. The magenta profile at $t=4.2 \times 10^{-1}$ s is the beginning of the wind phase.}
\label{fig:ExplosionDynamics}
\end{figure}

\begin{figure}
\centering{}
\includegraphics[width=0.9\linewidth]{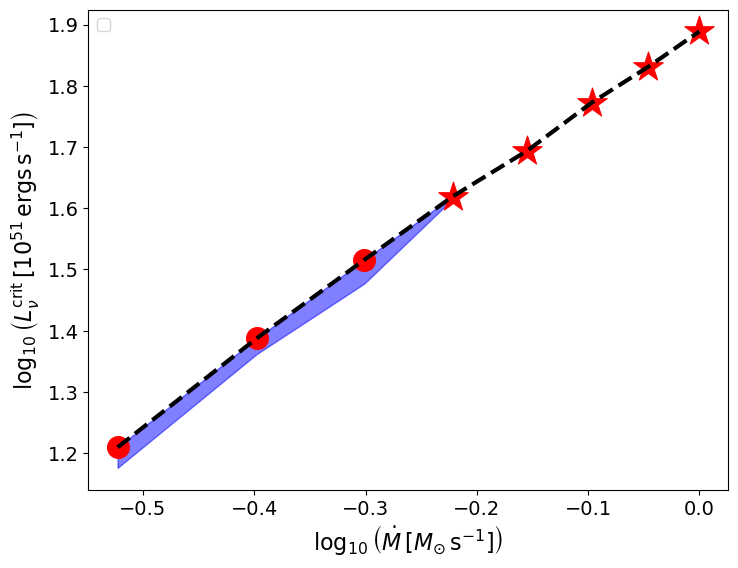}
\caption{Critical curve for $\mathcal{M}=2.0$ with fiducial inputs. Red dots represent models with strong shock oscillations that lead into explosions. Red stars are explosions that occur without oscillations. The blue shaded area shows regions with stable oscillatory models, while elsewhere under the curve is occupied by stable non-oscillatory models with the same fiducial inputs.}
\label{fig:CritCurveMachNumber}
\end{figure}

\begin{figure}
\centering{}
\includegraphics[width=0.9\linewidth]{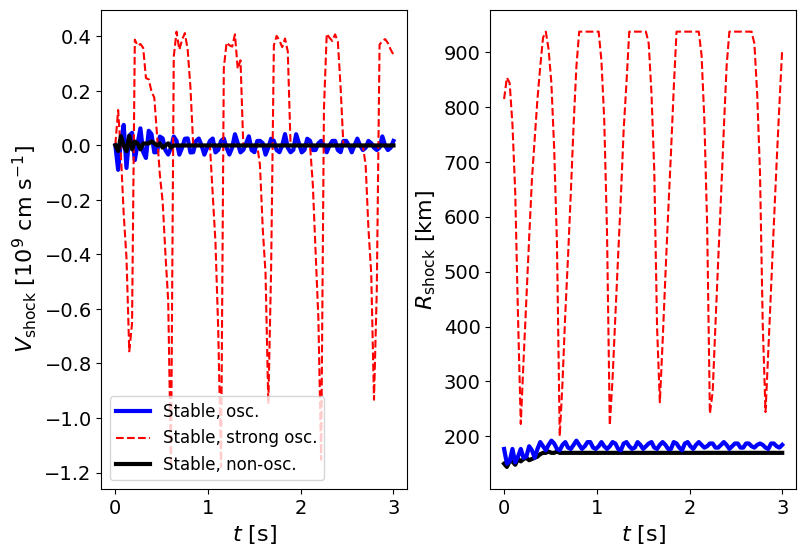}
\caption{Examples of stable and oscillatory accretion solutions. (Left) Shock velocities $V_{\mathrm{shock}}$ for models of varying time-dependent characteristics. All models here use $M_{\star}=1.4 \, \mathrm{M}_{\odot}$ and $\left< \epsilon_{\nu_{e}} \right> \simeq 12.6 \, \mathrm{MeV}$. The stable, non-oscillatory profile (solid black) uses $\mathcal{M}=1.3$, $\dot{M}=0.4 \, \mathrm{M}_{\odot} \, \mathrm{s}^{-1}$, $R_{\star}=30 \, \mathrm{km}$ and $L_{\nu}=17.0 \times 10^{51} \, \mathrm{ergs} \, \mathrm{s}^{-1}$. The stable, oscillatory profile (solid blue) uses $\mathcal{M}=1.3$, $\dot{M}=0.4 \, \mathrm{M}_{\odot} \, \mathrm{s}^{-1}$, $R_{\star}=30 \, \mathrm{km}$, and $L_{\nu}=17.4 \times 10^{51} \, \mathrm{ergs} \, \mathrm{s}^{-1}$. The stable model with strong oscillations (dashed red) uses $\mathcal{M}=2.0$, $\dot{M}=0.6 \, \mathrm{M}_{\odot} \, \mathrm{s}^{-1}$, $R_{\star}=50 \, \mathrm{km}$ and $L_{\nu}=10.4 \times 10^{51} \, \mathrm{ergs} \, \mathrm{s}^{-1}$. All models undergo an initial transient for the first few hundred ms, but the blue and red profiles continue oscillating for the duration of the simulation while the black model achieves perfect numerical steady state. (Right) Shock radii for the same simulations. $R_{\mathrm{shock}}$ stabilizes at $\sim 150-200$ km for the black and blue profiles, while the the red model oscillates at $\sim 600$ km.}
\label{fig:VshockRshock}
\end{figure}

\section{Results}
\label{sec:results1D}
Here we analyze the results from our parameterized 1D accretion models (see Section \ref{subsec:Fiducial} for a fiducial model set). We use these results both to examine the diagnostic tools that distinguish between stable and exploding models and to understand the importance of the input parameters in critical condition studies. We investigate the classic critical condition (Sections \ref{subsec:CritCurve} and \ref{subsubsec:CritCurveDepOnInputs}), the antesonic condition (Section \ref{subsec:Antesonic}), the heuristic heating and cooling timescale condition (Section \ref{subsec:advheat}), and the force explosion condition (Section \ref{subsec:FEC}). The results shown in Sections \ref{subsec:Fiducial}, \ref{subsec:Antesonic}, \ref{subsec:advheat}, and \ref{subsec:FEC} examine $\mathcal{M}$ dependence but otherwise use fiducial input parameters.  We also address the implications of the Si/O layer accreting onto the shock for the critical condition (Section \ref{subsec:MachNumberRealisticProgenitors}). 

\subsection{Fiducial models}
\label{subsec:Fiducial}
In Figure \ref{fig:VaryMachNumberMultiPanel}, we show time-steady fiducial solutions that share the same inputs, except we vary $\mathcal{M}=1.3, \, 2.0, \, 3.0$. We find that our accretion models and the critical condition changes with $\mathcal{M}$. All else equal, larger $\mathcal{M}$ drives down the pre-shock $P$ (equation \ref{eqn:PressureOuterboundary}), $T$, and $c_{s}$. Meanwhile, $P_{\mathrm{ram}}=\rho v_{r}^{2}$ ahead of the shock increases and drives the shock to smaller radii. This places the accretion solution in a more stable regime, which then requires larger $L_{\nu}^{\mathrm{crit}}$ to become unstable and explode. Note that $R_{\mathrm{shock}}$ decreases from $R_{\star}\sim 112 \, \mathrm{km}$ to $R_{\star} \sim 69 \, \mathrm{km}$ for $\mathcal{M}=1.3 \rightarrow 2.0$ and from $R_{\star}=69 \, \mathrm{km}$ to $R_{\star} \sim 64 \, \mathrm{km}$ for $\mathcal{M}=2.0 \rightarrow 3.0$, which shows that the flow ahead of the shock is approaching free-fall and further increasing $\mathcal{M}$ has less impact on the dynamics. We find that this convergence to pressureless free-fall manifests in all critical condition results at high $\mathcal{M}$ (see Sections \ref{subsubsec:CritCurveDepOnInputs}, \ref{subsec:Antesonic}, \ref{subsec:advheat}, and \ref{subsec:FEC}).

In Figure \ref{fig:PT2012_ReplicatePlot}, we increase $L_{\nu}$ over the range $[50, \, 77] \times 10^{51} \, \mathrm{ergs} \, \mathrm{s}^{-1}$ for a model using fiducial inputs with $\dot{M}=0.5 \, \mathrm{M}_{\odot} \, \mathrm{s}^{-1}$ and $\mathcal{M}=2.0$ until nearly achieving an explosion. As $L_{\nu}$ increases, this corresponds directly to an increase in $\dot{Q}$ (equation \ref{eqn:heatingrate}) and other thermal quantities in the post-shock region, and the shock moves toward larger radii. Here, the final stable model is at $L_{\nu}=77 \times 10^{51} \, \mathrm{ergs} \, \mathrm{s}^{-1}$, which is represented by the blue line in Figure \ref{fig:PT2012_ReplicatePlot}. Once $L_{\nu}^{\mathrm{crit}}=77.2 \times 10^{51} \, \mathrm{ergs} \, \mathrm{s}^{-1}$ is reached (Figure \ref{fig:ExplosionDynamics}), the flow destabilizes and the model explodes. The model transitions from accretion to explosion, where the shock travels outward and advects past $R_{\mathrm{out}}$. When this happens, $v_{r,\mathrm{out}}$ transitions from negative (inflow) to positive (outflow), which defines the explosion phase.

\subsection{Critical Curve}
\label{subsec:CritCurve}
For a fixed $\dot{M}$, $\mathcal{M}$, $M_{\star}$, $\left<\epsilon_{\nu_{e}}\right>$ and $R_{\star}$, we increase $L_{\nu}$ by $\Delta L_{\nu}=0.2 \times 10^{51}$ ergs s$^{-1}$ and evolve our accretion models until they achieve either non-oscillatory stability, oscillatory stability, or an explosion. When a model explodes, we set $L_{\nu}=L_{\nu}^{\mathrm{crit}}$, and we define near-critical $L_{\nu}^{\mathrm{crit,n}}$ models as those that are nearest-to-explosion and remain sub-critical. The critical values for the studies outlined in Sections \ref{subsec:Antesonic}, \ref{subsec:advheat}, and \ref{subsec:FEC}, are determined from $L_{\nu}^{\mathrm{crit,n}}$ models, because the calculations in these studies break down once stable accretion is no longer sustained. Otherwise, we use the first exploding model\footnote{Note that re-running the same models may produce different $L_{\nu}^{\mathrm{crit}}$ values at most by $\sim 1\%$.},  $L_{\nu}^{\mathrm{crit}}$, for the discussions in this section and Section \ref{subsubsec:CritCurveDepOnInputs}.

$L_{\nu}^{\mathrm{crit}}$ is shown in Figure \ref{fig:CritCurveMachNumber} for $\mathcal{M}=2.0$. In agreement with prior critical curve studies, we find that $L_{\nu}^{\mathrm{crit}}$ monotonically increases with $\dot{M}$. Similar to \cite{Gabey2015}, we observe stable oscillatory models just beneath $L_{\nu}^{\mathrm{crit}}$ for low $\dot{M}$. Figure \ref{fig:VshockRshock} shows the behavior of $R_{\mathrm{shock}}$ and $V_{\mathrm{shock}}$ for oscillating models. Note that the oscillations do not dampen during $t=3 \, \mathrm{s}$, and we observe strong shock oscillations for models with $R_{\star}=50 \, \mathrm{km}$. 

We find that the shock dynamics change as a function of $\mathcal{M}$. For $\mathcal{M}=1.3$, a stable oscillatory model occurs at high $\dot{M}$, specifically at $\dot{M}=1.0 \, \mathrm{M}_{\odot}$ with $L_{\nu}^{\mathrm{crit,n}}=49.8 \times 10^{51} \, \mathrm{ergs} \, \mathrm{s}^{-1}$. Meanwhile, no oscillations occur for models with $\dot{M}=0.3 \, \mathrm{M}_{\odot} \, \mathrm{s}^{-1}$ and $\mathcal{M}=3.0$. For some models with fiducial inputs, we also observe: (1) shock oscillations at $L_{\nu}^{\mathrm{crit}}$ but no oscillations for models at $L_{\nu}<L_{\nu}^{\mathrm{crit}}$ (e.g., models at $\dot{M}=0.5 \, \mathrm{M}_{\odot} \, \mathrm{s}^{-1}$ and $\mathcal{M}=1.3$), and (2) no shock oscillations at $L_{\nu}^{\mathrm{crit}}$, but shock oscillations for models at $L_{\nu}<L_{\nu}^{\mathrm{crit}}$ (e.g., models at $\dot{M}=0.6 \, \mathrm{M}_{\odot} \, \mathrm{s}^{-1}$ and $\mathcal{M}=2.0$).

\subsubsection{Critical Curve Dependence on Inputs}
\begin{figure*}
\centering{}
\includegraphics[width=0.8\linewidth]{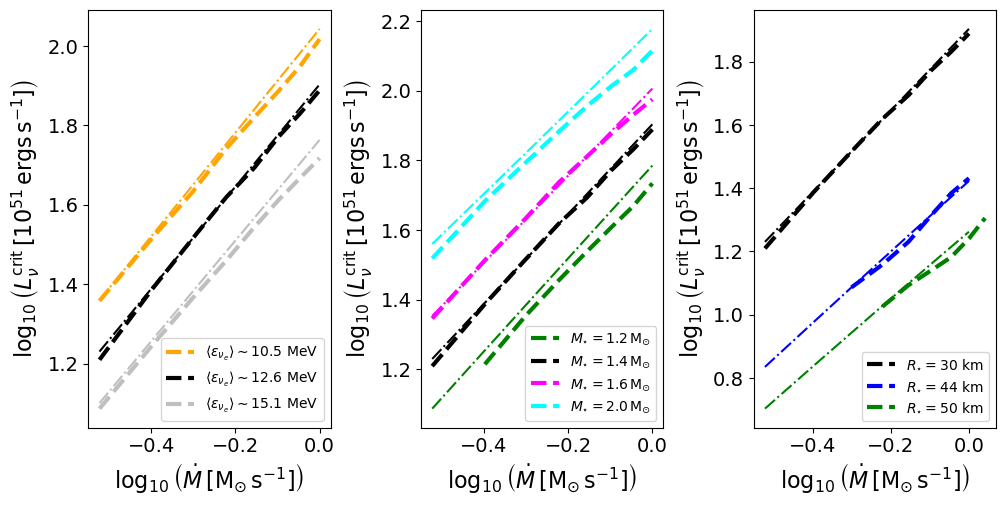}
\caption{$L_{\nu}^{\mathrm{crit}}$-$\dot{M}$ analysis for various input parameters at $\mathcal{M}=2.0$. (Left) $L_{\nu}^{\mathrm{crit}}$ at $R_{\star}=30 \, \mathrm{km}$ and $M_{\star}=1.4 \, \mathrm{M}_{\odot}$ with $\left< 
\epsilon_{\nu_{e}} \right> \simeq 10.5 \, \mathrm{MeV}$ (orange), $\left< 
\epsilon_{\nu_{e}}\right> \simeq 12.6 \, \mathrm{MeV}$ (black), and $\left< 
\epsilon_{\nu_{e}}\right> \simeq 15.1 \, \mathrm{MeV}$ (grey). (Middle) $L_{\nu}^{\mathrm{crit}}$ at $R_{\star}=30 \, \mathrm{km}$ and $\left< \epsilon_{\nu_{e}}\right> \simeq 12.6 \, \mathrm{MeV}$ with $M_{\star}=1.2 \, \mathrm{M}_{\odot}$ (green), $M_{\star}=1.4 \, \mathrm{M}_{\odot}$ (black), $M_{\star}=1.6 \, \mathrm{M}_{\odot}$ (magenta), and $M_{\star}=2.0 \, \mathrm{M}_{\odot}$ (cyan). (Right) $L_{\nu}^{\mathrm{crit}}$ at $M_{\star}=1.4 \, \mathrm{M}_{\odot}$ and $\left< \epsilon_{\nu_{e}}\right> \simeq 12.6 \, \mathrm{MeV}$ with $R_{\star}=30 \, \mathrm{km}$ (black), $R_{\star}=44 \, \mathrm{km}$ (blue), and $R_{\star}=50 \, \mathrm{km}$ (green). The thin dash-dotted lines with corresponding colors are the linear fittings provided by equation \ref{eqn:LnuCritDepOnInputs}.}
\label{fig:CriticalCurveInputAnalysis}
\end{figure*}

\begin{figure}
\centering{}
\includegraphics[width=0.9\linewidth]{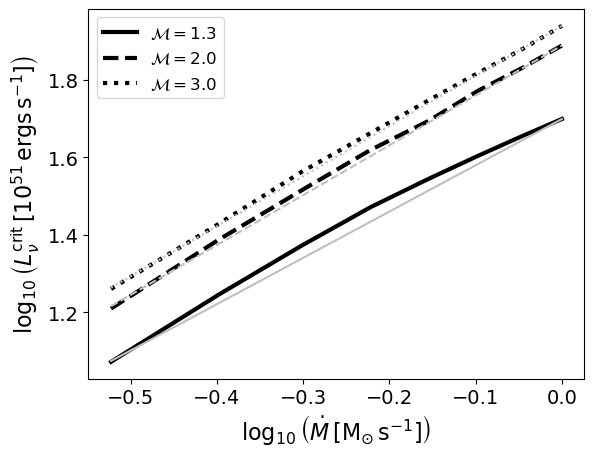}
\caption{Critical curves with fiducial inputs for $\mathcal{M}=1.3$ (solid), $\mathcal{M}=2.0$ (dashed), and $\mathcal{M}=3.0$ (dotted). The grey profiles of corresponding linestyle are the linear fittings provided by equation \ref{eqn:LnuCritDepOnInputs}. Each linear fitting is scaled with $L_{\nu}^{\mathrm{crit}}$ data at $\dot{M}=0.5 \, \mathrm{M}_{\odot} \, \mathrm{s}^{-1}$ from each value of $\mathcal{M}$.}
\label{fig:CritCurveMachStudyWithLinearFits}
\end{figure}

\begin{figure}
\centering{}
\includegraphics[width=0.9\linewidth]{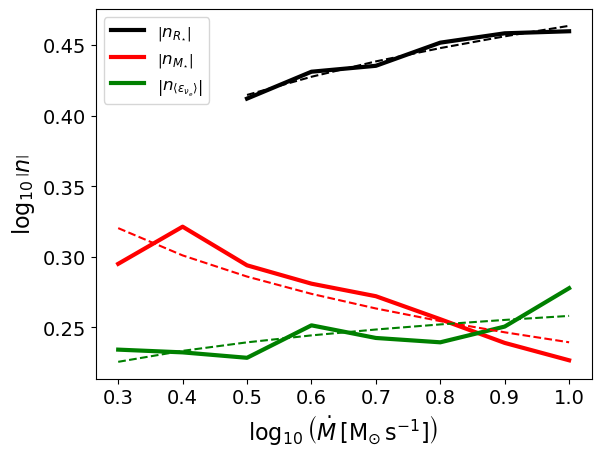}
\caption{Absolute value of $L_{\nu}^{\mathrm{crit}}$ power law scaling terms $n$ for $R_{\star}$ (solid black, equation \ref{eqn:nRstarPowerLawScaling}), $M_{\star}$ (solid red, equation \ref{eqn:nMstarPowerLawScaling}), and $\left< 
\epsilon_{\nu_{e}} \right>$ (solid green, equation \ref{eqn:nEnuePowerLawScaling}) as functions of $\dot{M}$. The dashed lines show the linear fits to the data.}
\label{fig:PowerLawVsMdot}
\end{figure}

We determine the power law scalings of $L_{\nu}^{\mathrm{crit}}$ with respect to the input parameters (shown in Figures \ref{fig:CriticalCurveInputAnalysis} and \ref{fig:CritCurveMachStudyWithLinearFits}) by using a least squares method to fit $L_{\nu}^{\mathrm{crit}}$ as a function of $\dot{M}$, $R_{\star}$, $M_{\star}$, and $\left<\epsilon_{\nu_{e}}\right>$, which yields the following scaling relation:
\label{subsubsec:CritCurveDepOnInputs}
\begin{equation}
\label{eqn:LnuCritDepOnInputs}
\begin{split}
    L_{\nu}^{\mathrm{crit}} = 32.8 \left(\frac{\dot{M}}{0.5 \, \mathrm{M}_{\odot} \, \mathrm{s}^{-1}}\right)^{1.29} \left(\frac{R_{\star}}{30 \, \mathrm{km}}\right)^{n_{R_{\star}}} \\ 
    \times \left(\frac{M_{\star}}{1.4 \, \mathrm{M}_{\odot}}\right)^{n_{M_{\star}}} \left(\frac{\left< \epsilon_{\nu_{e}} \right>}{12.6 \, \mathrm{MeV}}\right)^{n_{\epsilon}} \, 10^{51} \, \mathrm{ergs} \, \,  \mathrm{s}^{-1}.
\end{split}
\end{equation}
We find that the power laws weakly depend on $\dot{M}$, where,
\begin{equation}
\label{eqn:nRstarPowerLawScaling}
    n_{R_{\star}} = -2.58 \left(\frac{\dot{M}}{0.5 \,  \mathrm{M}_{\odot} \, \mathrm{s}^{-1}}\right)^{0.16},
\end{equation}
\begin{equation}
\label{eqn:nMstarPowerLawScaling}
    n_{M_{\star}} = 1.97 \left(\frac{\dot{M}}{0.5 \, \mathrm{M}_{\odot} \, \mathrm{s}^{-1}}\right)^{-0.15},
\end{equation}
and,
\begin{equation}
\label{eqn:nEnuePowerLawScaling}
    n_{\epsilon} = -1.69 \left(\frac{\dot{M}}{0.5 \, \mathrm{M}_{\odot} \, \mathrm{s}^{-1}}\right)^{0.06},
\end{equation}
Equations \ref{eqn:LnuCritDepOnInputs}-\ref{eqn:nEnuePowerLawScaling} are scaled with respect to one of the fiducial models at $\mathcal{M}=2.0$. Figure \ref{fig:PowerLawVsMdot} shows how these terms change with $\dot{M}$. The scaling term $n_{M_{\star}}$ decreases with $\dot{M}$, but the other terms positively trend with $\dot{M}$. We anticipate that these power laws may also depend on $R_{\star}$, $M_{\star}$, and $\left< \epsilon_{\nu} \right>$.

For the data in Figure \ref{fig:CriticalCurveInputAnalysis}, we select the following inputs: $\left< \epsilon_{\nu_{e}}\right> \in \left[10.5, \,  15.1\right] \, \mathrm{MeV}$ to capture the range of neutrino energies prevalent in CCSNe during accretion \citep{thompson2003, Scheck2006}, $M_{\star} \in \left[1.2, \, 2.0\right] \, \mathrm{M}_{\odot}$ to match the range of observations of NS masses \citep{Ferreira2021}, and $R_{\star} \in \left[30, \, 50\right] \, \mathrm{km}$ to capture a range of radii during the early accretion phase after collapse \citep{Burrows1986}. Figure \ref{fig:CriticalCurveInputAnalysis} shows how $L_{\nu}^{\mathrm{crit}}$ changes with the input parameters. Increasing $\left< \epsilon_{\nu_{e}} \right>$ causes $L_{\nu}^{\mathrm{crit}}$ to drop because $\left< \epsilon_{\nu_{e}}^{3} \right>/\left< \epsilon_{\nu_{e}} \right>$ enters in equation \ref{eqn:heatingrate} and directly controls the amount of neutrino-heating behind the shock. Higher $\left< \epsilon_{\nu_{e}} \right>$ leads to higher $\dot{Q}_{H}$ at fixed $L_{\nu}$, and thus lower $L_{\nu}^{\mathrm{crit}}$ is required for explosion. Increasing $M_{\star}$ causes $R_{\mathrm{shock}}$ to decrease, which leads to a higher post-shock $T$, a higher escape velocity, and more effective neutrino cooling. Higher $L_{\nu}^{\mathrm{crit}}$ is thus required to cause an explosion for higher $M_{\star}$. Finally, the heating rate drops with radius as $1/R_{\star}^2$ (equation \ref{eqn:heatingrate}). However, increasing $R_{\star}$ shifts the post-shock region to larger radii, which decreases $T$ behind the shock. $\dot{Q}_{c}$ decreases more rapidly with $R_{\star}$ than $\dot{Q}_{H}$ because $\dot{Q}_{C}$ sensitively depends on $T$ (equation \ref{eqn:coolingrate}). Because this drives the accretion flow closer to an explosive regime, $L_{\nu}^{\mathrm{crit}}$ decreases with larger $R_{\star}$.

As discussed in section \ref{subsec:Fiducial}, $L_{\nu}^{\mathrm{crit}}$ increases with the outer boundary  $\mathcal{M}$. Once $\mathcal{M}$ is sufficiently high ($\mathcal{M} \gtrsim 3.0$), the pre-shock accretion asymptotically approaches free-fall. Increasing $\mathcal{M}$ further does not qualitatively change the dynamics, and $L_{\nu}^{\mathrm{crit}}$ becomes unaffected. Our results show that $L_{\nu}^{\mathrm{crit}}$ increases on average by $\sim 44 \%$ and $\sim 12 \%$ for $\mathcal{M}=1.3 \rightarrow 2.0$ and $\mathcal{M}=2.0 \rightarrow 3.0$, respectively (see Figure \ref{fig:CritCurveMachStudyWithLinearFits} and Table \ref{table:MachNumberStudy}). We support this result by scaling equation \ref{eqn:LnuCritDepOnInputs} with $\mathcal{M}=1.3$ and $\mathcal{M}=3.0$ data, where Table \ref{table:MachNumberStudy} shows $L_{\nu}^{\mathrm{crit}}$ converging with respect to $\mathcal{M}$ for both equation \ref{eqn:LnuCritDepOnInputs} and the numerical data. Note that the decrease in the normalization of $L_{\nu}^{\mathrm{crit}}$ for $\mathcal{M}=2.0 \rightarrow 1.3$ is of order the decrease in $L_{\nu}^{\mathrm{crit}}$ due to the multi-dimensional instabilities studied in \cite{Murphy2008,Couch2013} ($\sim 30 \%$ decrease when comparing 1D data to 2D/3D data).

Note that \cite{Pejcha2012} have derived a scaling relation for $L_{\nu}^{\mathrm{crit}}$ with their fiducial model, where they find $L_{\nu}^{\mathrm{crit,PT}} \sim M_{\star}^{1.84} \, \dot{M}^{0.723} \,  R_{\star}^{-1.61}$. Because the power laws themselves depend on $\dot{M}$ (equations \ref{eqn:nRstarPowerLawScaling}-\ref{eqn:nEnuePowerLawScaling}),
we find that $L_{\nu}^{\mathrm{crit}}$ from equation \ref{eqn:LnuCritDepOnInputs} scales approximately the same as $L_{\nu}^{\mathrm{crit,PT}}$ with either $R_{\star}$ or $M_{\star}$ when either $\dot{M}=0.03 \, \mathrm{M}_{\odot} \, \mathrm{s}^{-1}$ or $\dot{M}=0.7 \, \mathrm{M}_{\odot} \, \mathrm{s}^{-1}$ respectively. Otherwise, we find that $L_{\nu}^{\mathrm{crit}}$ in our framework depends more sensitively on the inputs, which may be due to the following physical differences: (1) we assume $L_{\nu}$ is constant for all $r$, whereas \cite{Pejcha2012} include $L_{\nu}$ produced by the accreting matter, (2) in the pre-shock region, we set $P \neq 0$ with $\mathcal{M}$, while \cite{Pejcha2012} set $P=0$, and (3) our models are time-dependent, while the models by \cite{Pejcha2012} are time-independent. 

\cite{Pejcha2012} also derive a toy analytic condition on $L_{\nu}^{\mathrm{crit}}$. The toy model suggests that it scales linearly with $\dot{M}$ and $M_{\star}$ while having an inverse-square dependence on $R_{\star}$. Their model uses a simple relativistc EoS $\left(\epsilon = \frac{3P}{\rho}\right)$ with pressureless free-fall, basic scaling relations for neutrino heating and cooling, e.g, $H \sim 1/r^2$ and $C \sim 1/r^4$, and $\frac{dY_{e}}{dr}=0$. In our models, we implement a general EoS with finite pre-shock $P$ and a heating function that requires $Y_{e}\left(r\right)$ and $\left< \epsilon_{\nu_{e}} \right>$ as input. These physical differences change the normalization of $L_{\nu}^{\mathrm{crit}}$, which leads to the stronger scaling with the inputs shown in Figures \ref{fig:CriticalCurveInputAnalysis} and \ref{fig:CritCurveMachStudyWithLinearFits}. 

\begin{table*}
\begin{tabular}{ c||c|c|c|c|c|c|c|c|c| }
 \tikz{\node[below left, inner sep=8.0pt] (bottom) {$\mathcal{M}$};%
      \node[above right,inner sep=1.0pt] (top) {$\dot{M} \, [\mathrm{M}_{\odot} \, \mathrm{s}^{-1}]$};%
      \draw (bottom.north west|-top.north west) -- (bottom.south east-|top.south east);}
 & 0.3 & 0.4 & 0.5 & 0.6 & 0.7 & 0.8 & 0.9 & 1.0 & AVG\\
 \hline
 \hline
 $\mathcal{M}_{1}=1.3, \, \mathcal{M}_{2}=2.0$ (data) & 0.37 & 0.39 & 0.39 & 0.41 & 0.41 & 0.47 & 0.50 & 0.54 & 0.44 \\
 \hline
 $\mathcal{M}_{1}=2.0, \, \mathcal{M}_{2}=3.0$ (data) & 0.12 & 0.10 &  0.12 & 0.11 & 0.13 & 0.11 & 0.12 & 0.12 & 0.12 \\
 \hline
 \hline
 $\mathcal{M}_{1}=1.3, \, \mathcal{M}_{2}=2.0$ (eqn.) & 0.32 & 0.36 & 0.39 & 0.41 & 0.43 & 0.45 & 0.47 & 0.48 & 0.42 \\
 \hline
 $\mathcal{M}_{1}=2.0, \, \mathcal{M}_{2}=3.0$ (eqn.) & 0.11 & 0.11 & 0.12 & 0.12 & 0.12 & 0.12 & 0.12 & 0.12 & 0.12 \\
 \hline
\end{tabular}
\caption{Fractional increase in $L_{\nu}^{\mathrm{crit}}$, i.e., $\left|L_{\nu}^{\mathrm{crit}}\left(\mathcal{M}_{1}\right) - L_{\nu}^{\mathrm{crit}}\left(\mathcal{M}_{2}\right) \right| / L_{\nu}^{\mathrm{crit}}\left(\mathcal{M}_{1}\right)$, at each $\dot{M}$ when $\mathcal{M}$ is increased as $1.3 \rightarrow 2.0$ and $2.0 \rightarrow 3.0$. The 'data' rows use simulation data for $L_{\nu}^{\mathrm{crit}}$ with fiducial inputs. The 'eqn.' rows use output from equation \ref{eqn:LnuCritDepOnInputs}, where we use $L_{\nu}^{\mathrm{crit}}$ data at $\dot{M}=0.5 \, \mathrm{M}_{\odot} \, \mathrm{s}^{-1}$ for each $\mathcal{M}$ to scale equation \ref{eqn:LnuCritDepOnInputs}.}
\label{table:MachNumberStudy}
\end{table*}

\subsection{Antesonic Condition}
\label{subsec:Antesonic}
\cite{Pejcha2012,Raives2018,Raives2021} have shown that, for pressureless free-fall onto a standing accretion shock with an isothermal post-shock medium, there exists a critical condition on the post-shock isothermal sound speed $c_{T}$,
\begin{equation}
\label{eqn:antesonic}
\frac{\left(c_{T}^{\mathrm{crit}}\right)^2}{v_{\mathrm{esc}}^2} = \frac{3}{16}=0.1875,
\end{equation}
where $v_{\rm esc}$ is the escape velocity. Above this critical value of the ``antesonic ratio,"  there are no solutions that simultaneously satisfy the strong shock-jump conditions and the time-steady fluid equations in the post-shock flow for this model problem. The name ``antesonic" comes from the fact that this critical value is reached before the sonic condition in a trans-sonic isothermal Parker wind ($c_{s}^2/v_{\mathrm{esc}}^2=1/4$). \cite{Raives2018} showed that once the critical antesonic ratio in equation \ref{eqn:antesonic} is exceeded, the system undergoes a time-dependent transition from accretion to explosion and drives a thermal wind. Further, \cite{Raives2018} generalized the antesonic condition to a $\Gamma$ law EoS of the form $P\propto \rho^\Gamma$ and pressureless free-fall, obtaining a critical antesonic ratio of $c_{s}^2/v_{\mathrm{esc}}^2 = 3\Gamma/16$, which was obtained numerically by \cite{Pejcha2012}.

Because we are using a general EoS with finite pre-shock pressure, there likely does not exist an analytic derivation of the antesonic condition in our framework \citep{Raives2018}. In their steady-state calculations with a general EoS and neutrino heating/cooling, \cite{Pejcha2012} showed numerically that $\max(c_s^2/v_{\rm esc}^2)\simeq0.21$ is the maximum value of the antesonic condition, it occurs near $R_{\mathrm{gain}}$ in an individual profile, and it holds along the critical curve. 

\begin{figure*}
  \centering  \includegraphics[width=0.60\linewidth]{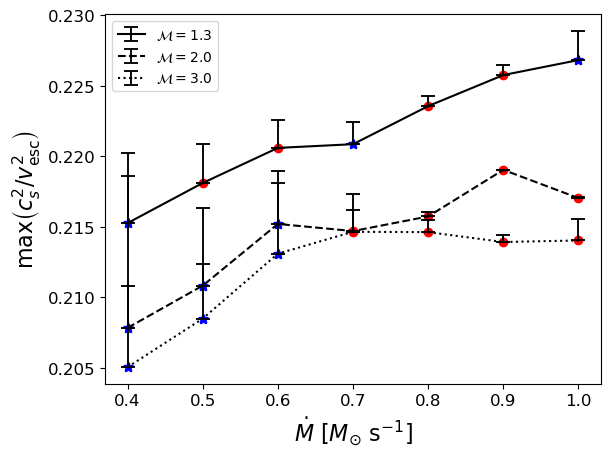}
  \caption{Maximum antesonic ratio for stable models with fiducial inputs at $L_{\nu}^{\mathrm{crit,n}}$ using $\mathcal{M}=1.3$ (solid line), $\mathcal{M}=2.0$ (dashed line) and $\mathcal{M}=3.0$ (dotted line). Blue stars represent oscillatory stable models, which use the time-averaged value of the maximum antesonic ratio. Red dots represent non-oscillatory stable models, which use the maximum antesonic ratio once steady-state is reached. The bars indicate the maximum value achieved over all $t$ during each simulation.} 
\label{fig:maxAntesonicVsMdot}
\end{figure*}

In order to assess the antesonic condition in our time-dependent simulations, we sample numerically stable models at $L_{\nu}^{\mathrm{crit,n}}$ for $\mathcal{M}=1.3$, $\mathcal{M}=2.0$, and $\mathcal{M}=3.0$ and we evaluate the maximum antesonic ratio for $R_{\star} \leq r \leq R_{\mathrm{shock}}$. For non-oscillatory stable models, we compute the maximum ratio once steady state is reached. For oscillatory stable models, we first evaluate the ratio at its post-shock maximum for all time $t$, then compute its average in $t$. We plot the maximum antesonic ratio for near-critical models in Figure \ref{fig:maxAntesonicVsMdot}. Note that the bars at each point indicate the true maximum value achieved for each simulation. This shows that time-dependent dynamics may temporarily drive the antesonic ratio above its critical value, but the model maintains stability.

\begin{figure}
\centering{}
\includegraphics[width=1.0\linewidth]{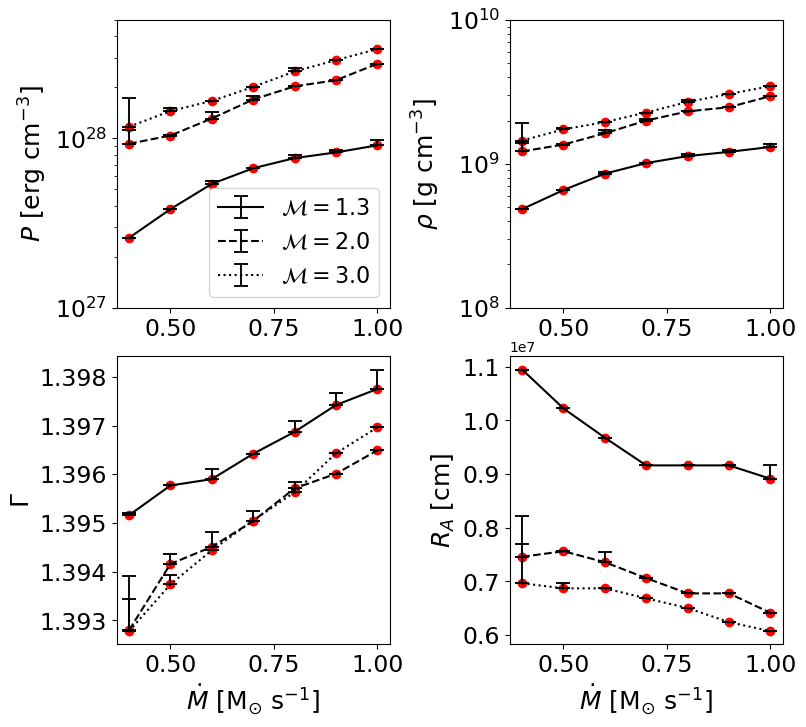}
\caption{Antesonic ratio quantities (equation \ref{eqn:AntesonicRatioSimplified}), i.e., $P$, $\rho$, $\Gamma$, and $R_{A}$ for near-critical fiducial models with $\mathcal{M}=1.3$ (solid), $\mathcal{M}=2.0$ (dashed), and $\mathcal{M}=3.0$ (dotted) plotted over $\dot{M}$. Similar to Figure \ref{fig:maxAntesonicVsMdot}, the bars indicate maximum values achieved during the simulation. The red dots are the representative values for each model, thus they are either the time-averaged value (for stable oscillatory models) or the final-time value (for stable non-oscillatory models).}
\label{fig:AntesonicErrorAnalysis3Mach}
\end{figure}

\cite{Pejcha2012} show that the critical antesonic ratio for models with accretion luminosity and a pre-shock free-fall velocity of $v_{\mathrm{ff}}=0.5 v_{\mathrm{esc}}$ remains roughly constant along the critical luminosity curve. However, we find that the value of $\mathrm{max}\left(c_{s}^2/v_{\mathrm{esc}}^2\right)$ increases as a function of $\dot{M}$ (and $L_{\nu}^{\mathrm{crit,n}}$) by $\sim 5 \%$ for $\mathcal{M}=1.3$, $\sim 3 \%$ for $\mathcal{M}=2.0$, and $\sim 4\%$ for $\mathcal{M}=3.0$, and that the near-critical ratio ranges between $\sim 0.205$ to $0.225$ (see Figure \ref{fig:maxAntesonicVsMdot}). To illustrate this, Figure \ref{fig:AntesonicErrorAnalysis3Mach} shows the relevant antesonic ratio quantities in the non-relativistic limit evaluated at the maximum antesonic ratio radius $R_{A}$,
\begin{equation}
\label{eqn:AntesonicRatioSimplified}
    \mathrm{max}\left(\frac{c_{s}^2}{v_{\mathrm{esc}}^2}\right)
    \simeq \frac{R_{A}}{2 G M}\frac{\Gamma P}{\rho},
\end{equation}
where their maxima in time are indicated by bars. $P$, $\rho$, and $\Gamma$ increase with $\dot{M}$ (and $L_{\nu}^{\mathrm{crit,n}}$), while $R_{A}$ decreases. Note that $P$ and $\rho$ increase while $\Gamma$ and $R_{A}$ decrease with $\mathcal{M}$, but all values begin converging for $\mathcal{M}\gtrsim 2.0$. The time-dependent changes in these quantities manifest as the changes shown in Figure \ref{fig:maxAntesonicVsMdot}.

\begin{figure}
  \centering  \includegraphics[width=0.9\linewidth]{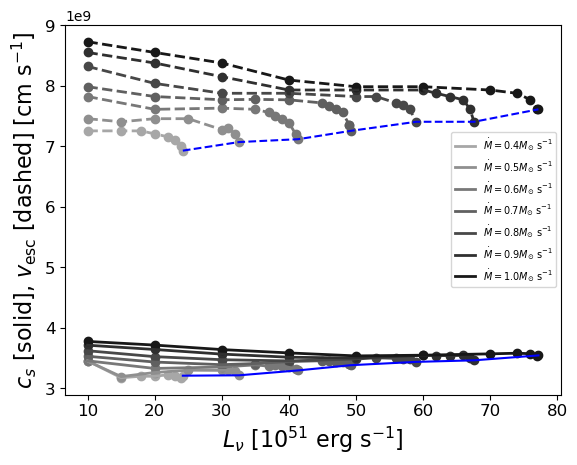}
  \includegraphics[width=0.9\linewidth]{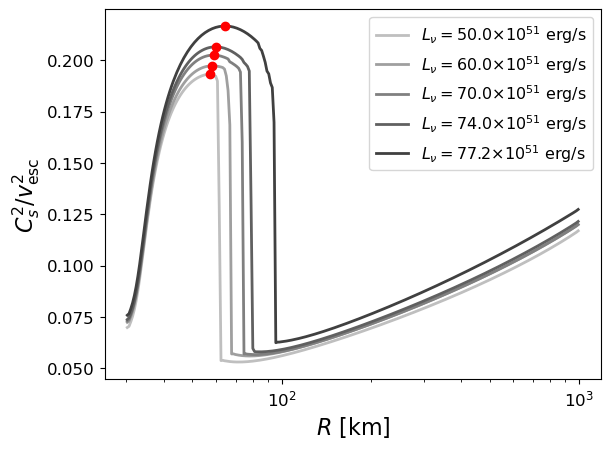}
  \caption{ (Top) Time averaged $c_{s}$ and $v_{\mathrm{esc}}$ as functions of $L_{\nu}$ and $\dot{M}$ evaluated at $R_{A}$ for $\mathcal{M}=2.0$. The dots show the evolution of $c_{s}$ and $v_{\mathrm{esc}}$ with $L_{\nu}$ at fixed $\dot{M}$, where $c_{s}$ and $v_{\mathrm{esc}}$ decrease in value and reach a minimum near the critical luminosity. As $\dot{M}$ is increased, this minimum is increased for both quantities, which is traced by a solid  blue line for $c_{s}$ and a dashed blue line for $v_{\mathrm{esc}}$. (Bottom) Time-averaged antesonic ratio for stable models beneath $L_{\nu}^{\mathrm{crit}}$ at $\dot{M}=1.0$ $M_{\odot}$ s$^{-1}$ and $\mathcal{M}=2.0$. The maximum of the ratio (red dot), which occurs around $R_{\mathrm{gain}}$, increases and is pushed to larger radii as criticality is neared.}
\label{fig:maxCsVescandMaxAnteVsRMach2pt0}
\end{figure}

When $L_{\nu}$ increases at fixed $\dot{M}$, $\rho$ increases more quickly than $\Gamma P$. As a result, $c_{s}$ is driven to smaller values and approaches a minimum at $L_{\nu}^{\mathrm{crit,n}}$ (see top panel of Figure \ref{fig:maxCsVescandMaxAnteVsRMach2pt0})
The bottom panel of Figure \ref{fig:maxCsVescandMaxAnteVsRMach2pt0} shows the time-averaged antesonic ratio as a function of $r$ over a range of $L_{\nu}$ at fixed $\dot{M}$. The maximum ratio, indicated by red dots, increases with $L_{\nu}$. Because $R_{A}$ increases with $L_{\nu}$ at fixed $\dot{M}$, this causes $v_{\mathrm{esc}}\left(R_{A}\right)$ to decrease (top panel of Figure \ref{fig:maxCsVescandMaxAnteVsRMach2pt0}). However, $v_{\mathrm{esc}}\left(R_{A}\right)$ decreases more rapidly with $L_{\nu}$ than $c_{s}\left(R_{A}\right)$, so the ratio $c_{s}^2/v_{\mathrm{esc}}^2$ increases with higher $L_{\nu}$. Because $L_{\nu}^{\mathrm{crit}}$ increases with $\dot{M}$, stables models at higher $\dot{M}$ have wider ranges of $c_{s}$ and $v_{\mathrm{esc}}$ values, which yields higher critical antesonic ratios. This is shown in Figure \ref{fig:maxAntesonicVsMdot}, where the maximum antesonic ratio generally increases with $L_{\nu}^{\mathrm{crit,n}}$ for $\mathcal{M}=1.3$ and $2.0$. For $\mathcal{M}=3.0$, the ratio increases with $\dot{M}$ for $\dot{M} \lesssim 0.6 \, \mathrm{M}_{\odot} \, \mathrm{s}^{-1}$ but  is roughly constant for greater $\dot{M}$ and exhibits the expected behavior for near-pressureless free-fall \citep{Pejcha2012}.

The ratio decreases with higher $\mathcal{M}$ for most models and begins converging after $\mathcal{M} \gtrsim 2.0$  (see Figure \ref{fig:maxAntesonicVsMdot}), e.g., it decreases by an average of $\sim 4\%$ for $\mathcal{M}=1.3 \rightarrow 2.0$ and $\sim 1\%$ for $\mathcal{M}=2.0 \rightarrow 3.0$. When the pre-shock $P$ drops ($\mathcal{M}$ is increased), $P_{\mathrm{ram}}$ at the shock increases (see Figure \ref{fig:VaryMachNumberMultiPanel}). This drives $R_{\mathrm{shock}}$ and  $R_{\mathrm{gain}}$ to lower radii. Lower $P$ leads to a subtle flattening of the post-shock $T$ gradient. Because $\dot{Q}_{C}$ sensitively depends on $T$, this subtle change in $T$ allows $Q_{H}$ to dominate $\dot{Q}_{C}$ at lower radii, thus leading to smaller $R_{\mathrm{gain}}$. As a result, $v_{\mathrm{esc}} \propto 1/\sqrt{R_{A}}$ increases with $\mathcal{M}$, where $R_{A} \sim R_{\mathrm{gain}}$ for non-isothermal models \citep{Pejcha2012,Raives2021}. Near $R_{\mathrm{gain}}$, $c_{s}$ also increases with $\mathcal{M}$. However, the absolute change in $v_{\mathrm{esc}}$ with respect to $\mathcal{M}$ is greater than the absolute change in $c_{s}$. This is what drives the near-critical antesonic ratio down in value for higher $\mathcal{M}$. 

The fiducial model set in \cite{Pejcha2012} predicts a near-constant maximum antesonic ratio along the critical curve. We find that, for a given set of inputs, there is always a critical value above which the flow transitions from accretion to explosion. However, while the changes across the range of $\dot{M}$ we explore are small, the critical antesonic ratio is not constant for low $\mathcal{M}$ and is a function of the inputs, e.g., $\dot{M}$, $\mathcal{M}$, and $L_{\nu}$, as shown in Figure \ref{fig:maxAntesonicVsMdot}. We find that pressurized pre-shock inflow modifies the antesonic condition, which is not captured by semi-analytic models that use $P=0$ as a boundary condition ahead of the shock. We also observe time-dependent dynamics that temporarily boost the maximum antesonic ratio  (Figures \ref{fig:maxAntesonicVsMdot} and \ref{fig:AntesonicErrorAnalysis3Mach}). Shock oscillations and pressurized free-fall are not accounted for in the models created by \cite{Pejcha2012}, therefore their models do not capture the full variation of $c_{s}^{2}/v_{\mathrm{esc}}^2$ along the critical curve. Nonetheless, the high-$\dot{M}$ and high-$\cal{M}$ models shown in Figure \ref{fig:maxAntesonicVsMdot} are remarkably consistent with the constant antesonic ratio determined by \cite{Pejcha2012}.

\subsection{Advection and Heating Timescales}
\label{subsec:advheat}
If the timescale for a fluid parcel to advect through the gain region $\tau_{\mathrm{adv}}$ is longer than the timescale for the gain region to be sufficiently heated via neutrino-matter interactions $\tau_{\mathrm{heat}}$, then the internal energy of the fluid inside the gain region will increase before the fluid accretes below $R_{\mathrm{gain}}$. This energy may be delivered to the shock and lead to an explosion, thus the comparison between these two timscales might provide a quantitative explosion condition \citep{ThompsonC2000,Thompson2005, Murphy2008}. Various definitions of the advection and heating timescales may be considered, which may lead to better quantitative descriptions of the critical luminosity. For example, the advection timescale may be defined as the time required for fluid to flow through a pressure scale height $H$ (e.g., \citealt{Thompson2005}):
\begin{equation}
\label{eqn:advectiontimescale}
    \tau_{\mathrm{adv},1}\left(r\right) = \left|\frac{H}{v_{r}} \right| = \left| \frac{\left(\frac{d \ln P}{dr}\right)^{-1}}{v_{r}}\right|,
\end{equation}
and the heating timescale may be defined as:
\begin{equation}
\label{eqn:heatingtimescale}
    \tau_{\mathrm{heat},1}\left(r\right) = \left|\frac{\epsilon}{\rho}\frac{1}{ \dot{Q}}\right| \sim \left|\frac{P}{\rho} \frac{1}{\dot{Q}}\right|,
\end{equation}
where $\dot{Q}$ is the net heating rate. Alternatively, the definition for the advection timescale could be formulated in terms of an integral to capture the total time in the gain region  \citep{Murphy2008},
\begin{equation}
\label{eqn:advectiontimescaleMB}
    \tau_{\mathrm{adv},2}\left(t\right)=\int_{R_{\mathrm{gain}}}^{R_{\mathrm{shock}}} \frac{dr}{\left|v_{r}\right|}.
\end{equation}
Similarly, one could define the heating timescale in spherical symmetry in terms of the ratio of the total internal energy in the gain region to the total net heating rate in the gain region \citep{Murphy2008,Pejcha2012,Murphy2017},
\begin{equation}
\label{eqn:heatingtimscaleMB}
    \tau_{\mathrm{heat},2}\left(t\right) =  \frac{ \int_{R_{\mathrm{gain}}}^{R_{\mathrm{shock}}} \, dr \, r^2 \, \rho \, \epsilon}{ \int_{R_{\mathrm{gain}}}^{R_{\mathrm{shock}}} \, dr\,  r^2 \, \rho \, \dot{Q} }.
\end{equation}
In these cases, one might expect the critical condition for explosion to be set by $\tau_{\rm adv}/\tau_{\rm heat}$ such that for
\begin{equation}
\label{eqn:advheatratio}
    \frac{\tau_{\mathrm{adv}}}{\tau_{\mathrm{heat}}} > 1.
\end{equation}
an explosion may soon follow.

\begin{figure*}
  \centering  \includegraphics[width=0.6\linewidth]{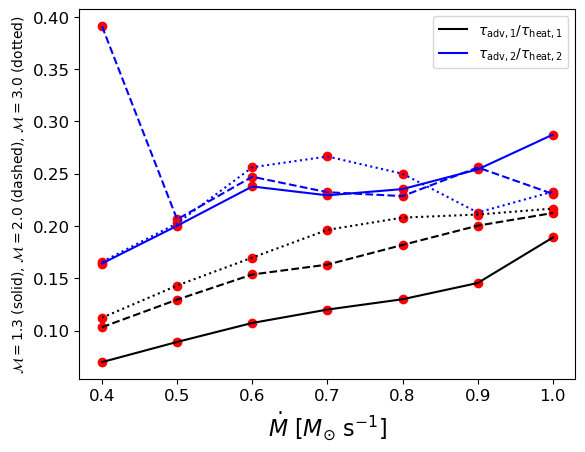}
  \caption{(Top) Maximum advection and heating timescale ratios for ${\mathcal{M}}=1.3$ (solid), $\mathcal{M}=2.0$ (dashed),  and $\mathcal{M}=3.0$ (dotted) for fiducial near-critical stable models. The black profile is calculated with equations \ref{eqn:advectiontimescale} and \ref{eqn:heatingtimescale}, and the blue profile is calculated with equations \ref{eqn:advectiontimescaleMB}. The ratios remain below unity until a model begins exploding. The black profiles demonstrate $\mathcal{M}$ dependence.}
  \label{fig:AllMaxTimescalesBoundaryValues}
\end{figure*}

\begin{figure}
  \centering  \includegraphics[width=0.9\linewidth]{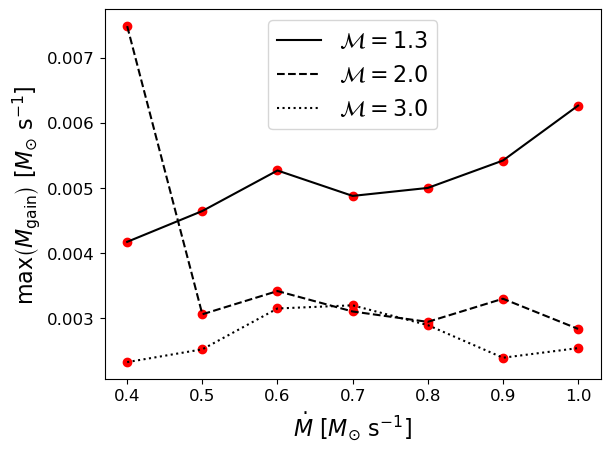}
  \caption{Maximum mass enclosed in the gain region $M_{\mathrm{gain}}$ for fiducial near-critical models. The behavior in $M_{\mathrm{gain}}$ profiles directly corresponds to the maximum timescale ratios with $\tau_{\mathrm{adv},2}$ (see Figure \ref{fig:AllMaxTimescalesBoundaryValues}). At $\dot{M}=0.4 \, \mathrm{M}_{\odot} \, \mathrm{s}^{-1}$ the $\mathcal{M}=2.0$ curve is driven up by shock oscillations.}
  \label{fig:maxTimescaleMgainBothMach}
\end{figure}

To test these timescale criteria, we evaluate $\mathrm{max}\left(\tau_{\mathrm{adv},1} / \tau_{\mathrm{heat},1}\right)$ and $\mathrm{max}\left(\tau_{\mathrm{adv},2} / \tau_{\mathrm{heat},2}\right)$ (see Figure \ref{fig:AllMaxTimescalesBoundaryValues}) at $L_{\nu}^{\mathrm{crit,n}}$. Note that we compute $\tau_{\mathrm{adv},1}$ and $\tau_{\mathrm{heat},1}$ with time-averaged quantities, then we find $\mathrm{max}\left(\tau_{\mathrm{adv},1}/\tau_{\mathrm{heat},1}\right)$ within the gain region. To understand how $\mathrm{max}\left(\tau_{\mathrm{adv},1}/\tau_{\mathrm{heat},1}\right)$ changes with $\mathcal{M}$, we consider the constituent terms in equations \ref{eqn:advectiontimescale} and \ref{eqn:heatingtimescale}. $R_{\mathrm{gain}}$ and $R_{\mathrm{shock}}$ decrease when $\mathcal{M}$ increases. As a result, the maximum timescale ratio is driven to smaller radii, i.e., regions with higher $P$ and $\epsilon$. $\dot{Q}$ in the gain region grows with larger $\mathcal{M}$, which contributes to lowering the timescale ratio. However, the changes in $P$, $r$, and $\epsilon$ overpower the changes in $\dot{Q}$ at the radius where the timescale ratio maximizes, and they all directly contribute to the growth of $\mathrm{max}\left(\tau_{\mathrm{adv},1}/\tau_{\mathrm{heat},1}\right)$ with $\mathcal{M}$. Similar to the antesonic ratio and $L_{\nu}^{\mathrm{crit}}$, $\mathrm{max}\left(\tau_{\mathrm{adv},1}/\tau_{\mathrm{heat},1}\right)$ converges as a function of $\mathcal{M}$ after $\mathcal{M} \gtrsim 2.0$ (see Figure \ref{fig:AllMaxTimescalesBoundaryValues}), e.g., the ratio increases by $\sim 42 \%$ for $\mathcal{M}=1.3 \rightarrow 2.0$ and $\sim 10 \%$ for $\mathcal{M}=2.0 \rightarrow 3.0$. Otherwise, $\mathrm{max}(\tau_{\mathrm{adv},1}/\tau_{\mathrm{heat},1})$ increases by a factor of two to three along $L_{\nu}^{\mathrm{crit,n}}$ depending on $\mathcal{M}$.

Excluding the data point at $\dot{M}=0.4 \, \mathrm{M}_{\odot} \, \mathrm{s}^{-1}$ and $\mathcal{M}=2.0$, we find that $\mathrm{max}\left(\tau_{\mathrm{adv},2}/\tau_{\mathrm{heat},2}\right)$ increases by a factor of $0.5$ to $0.7$ depending on $\mathcal{M}$. This is due to the changes in the post-shock $\rho$ in response to $\dot{M}$ and $L_{\nu}$, which incurs changes in $M_{\mathrm{gain}}$ that directly map into the maximum timescale ratio (see Figure \ref{fig:maxTimescaleMgainBothMach}). This timescale ratio does not have a strong dependence on $\mathcal{M}$ compared to $\mathrm{max}\left(\tau_{\mathrm{adv},1}/\tau_{\mathrm{heat},1}\right)$. In contrast to the results in \cite{Pejcha2012}, we find that $\mathrm{max}\left(\tau_{\mathrm{adv,2}} / \tau_{\mathrm{heat,2}}\right)$ does not monotonically increase with $L_{\nu}^{\mathrm{crit,n}}$ (see Figure \ref{fig:AllMaxTimescalesBoundaryValues}).

\begin{figure}
\centering{}
\includegraphics[width=0.9\linewidth]{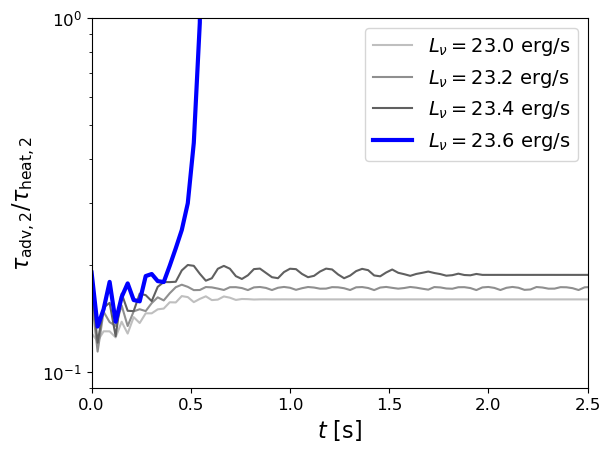}
\caption{Advection and heating timescale ratio for equations \ref{eqn:advectiontimescaleMB} and \ref{eqn:heatingtimscaleMB} plotted over $t$ for fiducial models with ${\dot{M}=0.5}$ $M_{\odot}$ s$^{-1}$ and $\mathcal{M}=1.3$. Models with $L_{\nu}<23.6 \times 10^{51}$ ergs s$^{-1}$ are stable, whereas $L_{\nu}^{\mathrm{crit}}=23.6 \times 10^{51}$ ergs s$^{-1}$ (blue line) explodes. At $L_{\nu}^{\mathrm{crit}}$, the calculation stops once the shock reached the outer boundary and the ratio exceeds unity. Oscillations in the profiles are due to oscillations in the shock as criticality is approached. Note that $\mathrm{max}\left(\tau_{\mathrm{adv},1} / \tau_{\mathrm{heat},1}\right)$ demonstrates similar behavior.} 
\label{fig:TimescaleRatioMach1pt3Mdot0pt3}
\end{figure}

Our sub-critical models maintain $\tau_{\mathrm{adv}}/\tau_{\mathrm{heat}}<1$. Some models, e.g., the model with $\mathcal{M}=2.0$ and $\dot{M}=0.4 \, \mathrm{M}_{\odot} \, \mathrm{s}^{-1}$ in Figure \ref{fig:AllMaxTimescalesBoundaryValues}, may have larger timescale values than others due to shock oscillations, which extends the gain region and may increase the value of $M_{\mathrm{gain}}$ or alter the post-shock $P$, $\epsilon$, and $\dot{Q}$ values. Otherwise, the timescale ratio for near-critical models is similar to the 1D values obtained by \cite{Murphy2008}. We confirm that $\tau_{\mathrm{adv}}/\tau_{\mathrm{heat}}$ exceeds unity after the explosion sets in. This is demonstrated in Figure \ref{fig:TimescaleRatioMach1pt3Mdot0pt3}, which shows the time evolution of $\tau_{\mathrm{adv},2}/\tau_{\mathrm{heat},2}$ for models with successively higher $L_{\nu}$.

\subsection{Force explosion condition (FEC)}
\label{subsec:FEC}
\cite{Murphy2017, Gogilashvili2022, Gogilashvili2023} have shown that by integrating the equations for fluid dynamics, they arrive at an expression that computes the net forces acting on the shock. Using this, they show that all models above the critical curve proposed in \cite{Burrows1993} have a force imbalance for which no stable solutions exist and $V_{\mathrm{shock}}$ must be positive. To derive this condition, \cite{Murphy2017, Gogilashvili2022} integrate in volume the momentum equation (equation \ref{eqn:MomentumConservation}) for time-independent, spherically symmetric accretion flow, which leads to an equation for the net forces acting on the shock. The sum of these forces gives the integral FEC $\Psi$:

\begin{equation}
\label{eqn:ForceExplosionCondition}
\begin{split}
\Psi=\left(P_{\star} + \rho_{\star} v_{r,\star}^{2}\right)R_{\star}^{2} - \left(P_{s+} + \rho_{s+} v_{r,s+}^{2}\right)R_{s}^2 \\
+ \int_{R_{\star}}^{R_{s}} 2Pr \, dr - GM_{\star} \int_{R_{\star}}^{R_{s}} \rho \, dr
\end{split}
\end{equation}
where ``$s+$" indicates the region just above the shock. Equation \ref{eqn:ForceExplosionCondition} computes the net forces on the shock due to $P_{\mathrm{ram}}$, the pressure gradient, and gravity. The FEC is typically evaluated in the dimensionless form $\tilde{\Psi}$ \citep{Gogilashvili2022},
\begin{equation}
\label{eqn:dimensionlessFEC} 
    \tilde{\Psi} = \frac{{\Psi}}{\dot{M}_{1.0} \sqrt{\frac{G M_{\star}}{R_{\star}}}  \, \mathrm{M}_{\odot} \, \mathrm{s}^{-1}},
\end{equation}
where we choose to normalize ${\Psi}$ by $\dot{M}_{1.0}=\dot{M}/1.0 \, \mathrm{M}_{\odot} \, \mathrm{s}^{-1}$. \cite{Murphy2017,Gogilashvili2022} show that one can evaluate $\tilde{\Psi}$ as a function of $x_{\mathrm{shock}}=R_{\mathrm{shock}}/R_{\star}$ by solving the time-independent Euler equations with the strong shock jump conditions. If $\mathrm{min}(\tilde{\Psi}(x_{\mathrm{shock}}))<0$, then $\tilde{\Psi}( 
x_{\mathrm{shock}})$ crosses an equilibrium state where $V_{\mathrm{shock}}=0$. In this case, $\tilde{\Psi}>0$ or $\tilde{\Psi}<0$ corresponds to either $V_{\mathrm{shock}}>0$ or $V_{\mathrm{shock}}<0$ respectively, which indicates that the shock may be forced toward the equilibrium state. Otherwise, if $\mathrm{min}( \tilde{\Psi} (x_{\mathrm{shock}}) )>0$, then $V_{\mathrm{shock}}>0$ for all $x_{\mathrm{shock}}$, and no stable solution exists. Further, \citep{Murphy2017} performed this analysis by using inputs from their numerical models at each time step to determine $\mathrm{min}(\tilde{\Psi}(x_{\mathrm{shock}}))$, which showed how close their accretion models were to exploding at each moment, i.e., a ``nearness-to-explosion" condition. Alternatively, a fully analytic FEC can be derived from $\mathrm{min}(\tilde{\Psi})=0$, which has been shown to accurately predict explosions for numerical simulations \citep{Gogilashvili2022,Gogilashvili2023}.

\begin{figure}
  \centering  \includegraphics[width=0.9\linewidth]{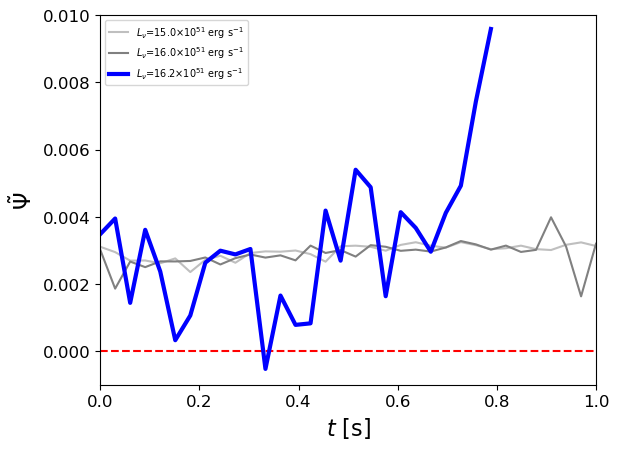}
  \includegraphics[width=0.9\linewidth]{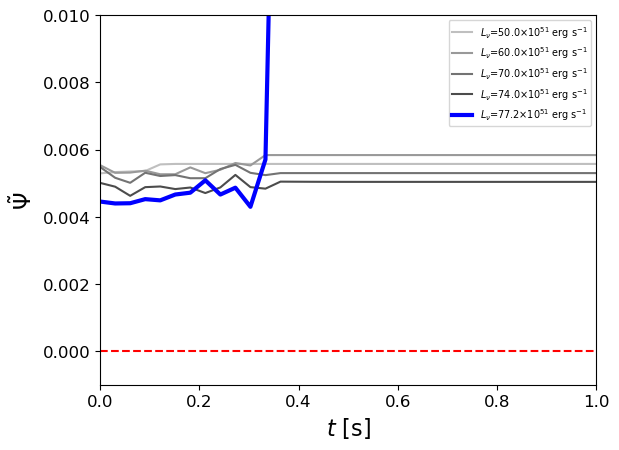}
  \caption{$\tilde{\Psi}$ as a function of time. (Top) $\tilde{\Psi}$ for a series of models at $\dot{M}=0.3 \, \mathrm{M}_{\odot} \, \mathrm{s}^{-1}$ and $\mathcal{M}=2.0$ with fiducial inputs. $\tilde{\Psi}$ oscillates around $3.0 \times 10^{-3}$ for stable models. At $L_{\nu}^{\mathrm{crit}}=16.2 \times 10^{51} \, \mathrm{ergs} \, \mathrm{s}^{-1}$ (blue), $\tilde{\Psi}$ is driven up in value. The red dashed line at $\tilde{\Psi}=0$ shows the expected value of $\tilde{\Psi}$ for non-oscillatory stable models. (Bottom) Same configuration as the top panel, but for models at $\dot{M}=1.0 \, \mathrm{M}_{\odot} \, \mathrm{s}^{-1}$ with $L_{\nu}^{\mathrm{crit}}=77.2 \times 10^{51} \, \mathrm{ergs} \, \mathrm{s}^{-1}$. All sub-critical models here become time-steady for $t>0.35 \, \mathrm{s}$.}
\label{fig:PsiVst}
\end{figure}

To assess the FEC, we compute $\tilde{\Psi}$ directly from equation \ref{eqn:dimensionlessFEC} using simulation data. If $\tilde{\Psi}(t) \sim 0$, this indicates that the shock is stable. Otherwise, $\tilde{\Psi}(t)>0$ indicates an explosion. Unlike the nearness-to-explosion condition, this procedure does not indicate how close a model is to exploding because $\tilde{\Psi}(t)$ is not necessarily being evaluated at its minimum with respect to $R_{\mathrm{shock}}$. However, it can still be used to identify an explosion once it occurs \citep{Murphy2017}. Figure \ref{fig:PsiVst} shows the time evolution of $\tilde{\Psi}$ for several stable models and an exploding model (blue). On average, we find positive values of $\tilde{\Psi}$ for stable models, but we also find that shock oscillations can briefly drive $\tilde{\Psi}$ down to negative values (top panel of Figure \ref{fig:PsiVst}).
Once $L_{\nu}^{\mathrm{crit}}$ is exceeded and the shock moves out,  $\tilde{\Psi}$ rapidly increases. In this manner, equation \ref{eqn:dimensionlessFEC} operates similarly to tracking $R_{\mathrm{shock}}$ in time \citep{Murphy2017}.

\begin{figure}
\centering{}
\includegraphics[width=0.9\linewidth]{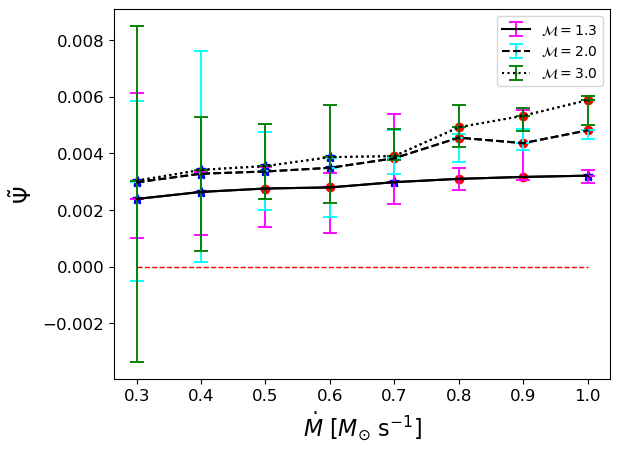}
\caption{$\tilde{\Psi}$ for near-critical models with fiducial inputs as a function of $\dot{M}$ at $\mathcal{M}=1.3$ (solid, magenta cap), $\mathcal{M}=2.0$ (dashed, cyan cap), and $\mathcal{M}=3.0$ (dotted, green cap). Blue stars indicate oscillatory stable models, which use the time-averaged value of $\tilde{\Psi}$. Red dots indicate non-oscillatory stable models, which use the value of $\tilde{\Psi}$ at the final timestep. The bars indicate the minimum and maximum variation of $\tilde{\Psi}$ in time. Stable models should have $\tilde{\Psi}=0$, but we find $\tilde{\Psi}$ to generally be slightly positive.}
\label{fig:NearCriticalPsi}
\end{figure}

In Figure \ref{fig:NearCriticalPsi}, we show summary results along the critical curve. Similar to \cite{Murphy2017}, we find that $\tilde{\Psi} \gtrsim 0$ for near-critical models. Especially at low $\dot{M}$, time-dependent fluctuations may drive $\tilde{\Psi}$ above and below zero (Figure \ref{fig:NearCriticalPsi}). We also find that $\tilde{\Psi}$ increases by a factor of $\simeq 0.2-2$ along $L_{\nu}^{\mathrm{crit,n}}$ depending on $\mathcal{M}$. This behavior is linked to the properties of the post-shock structure. More matter accretes through the shock at higher $\dot{M}$, which increases the post-shock thermal content and densities. The integrated pressure and density terms in equation \ref{eqn:ForceExplosionCondition} increase with $\dot{M}$ in response to these changes in the post-shock flow, but the relative difference between these terms decreases with $\dot{M}$. This causes $\tilde{\Psi}$ to increase along $\dot{M}$ for near-critical models. Note that in equation \ref{eqn:ForceExplosionCondition}, the pre-shock quantities are approximately an order of magnitude less than all the other terms, thus their response to changes in $\dot{M}$ are negligible. Meanwhile, the quantities located at $R_{\star}$ are approximately constant for all $\dot{M}$.

$\tilde{\Psi}$ depends on $\mathcal{M}$ (Figure \ref{fig:NearCriticalPsi}), but converges as $\mathcal{M}$ is increased. We observe an average change of $\sim 24 \%$ in $\tilde{\Psi}$ when $\mathcal{M}=1.3 \rightarrow 2.0$ and $\sim 8 \%$ when $\mathcal{M}=2.0 \rightarrow 3.0$. As shown in Figure \ref{fig:VaryMachNumberMultiPanel}, changing $\mathcal{M}$ modifies the position of $R_{\mathrm{shock}}$, which maps to equation \ref{eqn:ForceExplosionCondition} through the limits of integration. When $R_{\mathrm{shock}}$ is driven to smaller radii through $\mathcal{M}$, both integrated quantities in equation \ref{eqn:ForceExplosionCondition} decrease. Because the integral over $G M_{\star} \rho$ dominates in $\tilde{\Psi}$, the overall value of $\tilde{\Psi}$ increases when the integrand $G M_{\star} \rho$ decreases. As such, $\tilde{\Psi}$ depends on $\mathcal{M}$ primarily through $R_{\mathrm{shock}}$.

\subsection{Implications of $\mathcal{M}$ for realistic progenitors}
\label{subsec:MachNumberRealisticProgenitors}
Previous work implies a direct connection between the accretion of compositional interfaces and the initiation of explosion \citep{Murphy2008,hanke2013,suwa2015,Summa2018,Ott2018a, vartanyan2018, Wang2022,Burrows2024}. Specifically, the rapid decrease in density across the Si/O interface is important for explosion because once it accretes onto the stalled accretion shock, it causes a decrease in both $\dot{M}$ and $P_{\mathrm{ram}}$ \citep{vartanyan2018}. Note that the timing of the accretion of the Si/O interface and the magnitude of the decrease in $\dot{M}$ is progenitor-dependent (e.g., \citealt{Woosley2002,Woosley2007,sukhbold2016,sukhbold2018}). 

\begin{figure}
  \centering
  \includegraphics[width=0.9\linewidth]{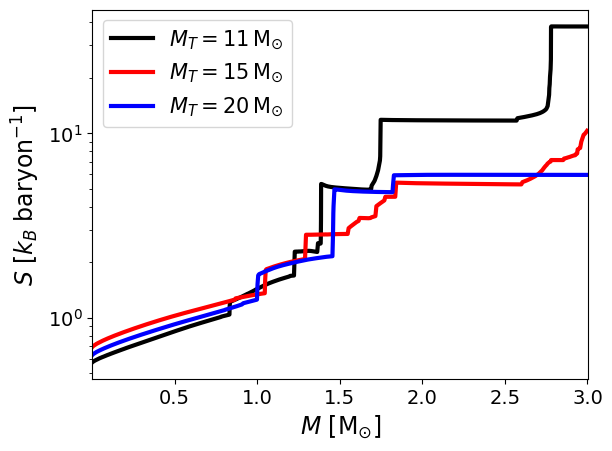}
  \includegraphics[width=0.9\linewidth]{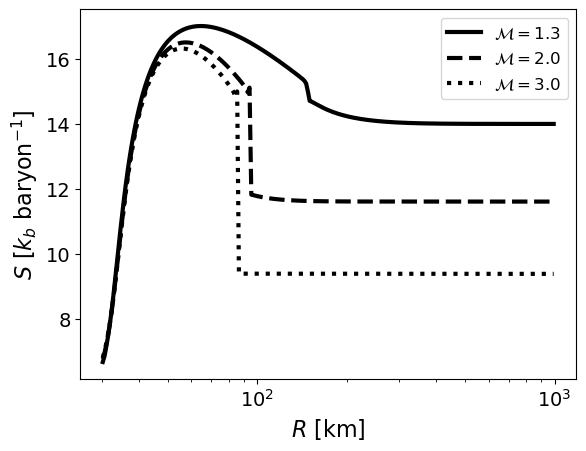}
  \caption{(Top) $S$ profiles (in units of $k_{b} \, \mathrm{baryon}^{-1}$) as functions of mass (in units of $M_{\odot}$) for pre-collapse Solar metallicity progenitors of $M_{T}=11 \, \mathrm{M}_{\odot}$ (black), $M_{T}=15 \, \mathrm{M}_{\odot}$ (red), and $M_{T}=20 \, \mathrm{M}_{\odot}$ (blue) from the \protect\cite{Woosley2007} progenitor data sets. The first steep gradient jump in the profiles all located at $\sim 1-2 \, \mathrm{M}_{\odot}$ corresponds to the Si/O interface. (Bottom) Entropy profiles for fiducial models near-criticality at $\dot{M}=1.0 \, \mathrm{M}_{\odot} \, \mathrm{s}^{-1}$. $S$ is approximately constant for a given $\mathcal{M}$ at $R \sim 1000 \, \mathrm{km}$ and increases by $\sim 18\%$ at $R \sim 1000 \, \mathrm{km}$ whenever $\Delta \mathcal{M} \sim +1.0$. The $\mathcal{M}=1.3$ profile has a non-zero gradient in the pre-shock region because $\dot{Q}$ is inadequately suppressed by equation \ref{eqn:NucleonMassFraction} in this parameter space ($\mathcal{M}=1.3$, $\dot{M}=1.0 \, \mathrm{M}_{\odot} \, \mathrm{s}^{-1}$ and $L_{\nu} \sim 50 \times 10^{51} \, \mathrm{ergs} \, \mathrm{s}^{-1}$.}
\label{fig:EntropyProfilesProgenitors}
\end{figure}

In the context of the critical curve, it is natural to associate a decrease in $\dot{M}$ at approximately fixed $L_{\nu}$ as the condition needed to cross the critical threshold for explosion (e.g., Figures \ref{fig:CriticalCurveInputAnalysis} and \ref{fig:CritCurveMachStudyWithLinearFits}). However, we highlight an additional connection between the accretion of the Si/O layer and explosion. Our results in Figure \ref{fig:CritCurveMachStudyWithLinearFits} show that, at fixed $\dot{M}$, the normalization of the critical curve depends on the thermal content of the accreted matter, as parameterized by $\mathcal{M}$. The top panel of Figure \ref{fig:EntropyProfilesProgenitors} shows multiple $S$ profiles from the \cite{Woosley2007} data set as functions of the mass coordinate $M$. Each discontinuity is associated with a compositional interface, where the Si/O layer is located at $ \sim M=1.25 \, \mathrm{M}_{\odot}$ for the total mass $M_{T}=11 \, \mathrm{M}_{\odot }$ model, $\sim M=1.7 \, \mathrm{M}_{\odot}$ for the $M_{T}=15 \, \mathrm{M}_{\odot}$ model, and $\sim 1.7 \, \mathrm{M}_{\odot}$ for the $M_{T}=20 \, \mathrm{M}_{\odot}$ model. For a given mass shell, the entropy profiles are roughly constant, and they only change drastically across a composition interface. When these interfaces meet the shock, they drive the accretion flow closer to explosion in two ways simultaneously: 1) $\dot{M}$ directly decreases because $\rho$ decreases, and 2) $L_{\nu}^{\mathrm{crit}}$ decreases because higher $S$, i.e., lower $\mathcal{M}$, material accretes onto the shock (e.g., Figure \ref{fig:CritCurveMachStudyWithLinearFits}). Given our discussion here and in Section \ref{subsubsec:CritCurveDepOnInputs}, we argue that the changes in the thermal content (e.g., $S$ or $\mathcal{M}$) across a compositional interface are important when considering the explodability of massive star progenitors. The increase in the entropy of the accreted material leads to lower critical neutrino luminosity, thus the change in the thermal content of the accreted material across compositional interfaces must be considered along with the changes in $\dot{M}$.

\section{Conclusions}
\label{sec:conclusions}
Using Athena++, we have constructed core-collapse supernova accretion models in time-dependent spherical symmetry with a general EoS and optically thin neutrino heating and cooling to more fully understand the critical neutrino luminosity $L_{\nu}^{\mathrm{crit}}$ and to analyze the explosion conditions, e.g., the antesonic condition, the heuristic timescale condition, and the force explosion condition (FEC). Our input parameters include the neutrino luminosity $L_{\nu}$, average neutrino energy $\left<\epsilon_{\nu}\right>$, accretion rate $\dot{M}$, Mach number of the accreted material $\mathcal{M}$, proto-neutron star radius $R_{\star}$, mass $M_{\star}$, and neutrino optical depth $\tau$. We use $L_{\nu}$ and $\dot{M}$ to map the critical curve, and we study the explosion  conditions (Sections \ref{subsec:Antesonic}, \ref{subsec:advheat}, and \ref{subsec:FEC}) using the steady accretion solutions nearest to explosion $L_{\nu}^{\mathrm{crit,n}}$. We evolve models up to $t=3 \, \mathrm{s}$ to achieve non-oscillatory stability (predominantly at high $\dot{M}$), oscillatory stability (predominantly at low $\dot{M}$), or explosion. We summarize our findings as follows:
\begin{itemize}
  \item Similar to past studies, we quantify how the critical neutrino luminosity $L_{\nu}^{\mathrm{crit}}$ for explosion depends on the key parameters of the problem, including the PNS radius, mass, and the average neutrino energy (see Section \ref{subsubsec:CritCurveDepOnInputs} and Figure \ref{fig:CriticalCurveInputAnalysis}). In particular, we find the numerical results $L_{\nu}^{\mathrm{crit}} \propto R_{\star}^{-2.58}$,  $L_{\nu}^{\mathrm{crit}} \propto M_{\star}^{1.97}$, $L_{\nu}^{\mathrm{crit}} \propto \left< \epsilon_{\nu_{e}}\right>^{-1.69}$, and $L_{\nu}^{\mathrm{crit}} \propto \dot{M}^{1.29}$. We generally find time-dependent oscillatory solutions (see \citealt{Gabey2015}) at low $\dot{M}$ and time-steady non-oscillatory accretion at high $\dot{M}$.
  
  \item  Different from previous works, we study the dependence of $L_{\nu}^{\mathrm{crit}}$ on the thermal content of the accreted matter, as parameterized by the gas Mach number ($\cal{M}$) at the outer edge of our computational domain. Higher $\cal{M}$ corresponds to lower thermal content,  closer to the limit of pressure-less free-fall. Importantly, we show that  $L_{\nu}^{\mathrm{crit}}$ decreases as the thermal content of the accreted material increases ($\cal{M}$ decreases; see Section \ref{subsubsec:CritCurveDepOnInputs}). For fixed $\dot{M}$, lower $\mathcal{M}$ decreases the ram pressure at the shock (see Figure \ref{fig:VaryMachNumberMultiPanel}), leading to lower  $L_{\nu}^{\mathrm{crit}}$ for explosion. As expected, the value of  $L_{\nu}^{\mathrm{crit}}$ converges as $\mathcal{M}$ increases because the system approaches the limit of pressureless free-fall (high $\mathcal{M}$; see Figure \ref{fig:CritCurveMachStudyWithLinearFits}). The fractional decrease in the normalization of the critical curve is of order the decrease from multi-dimensional instabilities highlighted in past works on the supernova mechanism \citep{Murphy2008, Couch2013}.
  
  \item Because of the influence of the thermal content on the normalization of the critical curve, we argue in Section \ref{subsec:MachNumberRealisticProgenitors} that the accretion of compositional interfaces onto a stalled accretion shock provides two (not one) effects that contribute to the possibility of an explosion: (1) as is well know, $\dot{M}$ decreases  because of the rapid decrease in density at compositional interfaces in a massive star progenitor, and (2), as we show here, the increase in entropy across a compositional interface in a massive star (see Figure \ref{fig:EntropyProfilesProgenitors}) leads to a lower overall normalization of $L_\nu^{\rm crit}$ at any $\dot{M}$ (see Figure \ref{fig:CritCurveMachStudyWithLinearFits}). 
  
  \item We assess several critical conditions for explosion that have been suggested in the literature, including the so-called ``antesonic" condition (see Section \ref{subsec:Antesonic}). We find a nearly-constant critical antesonic ratio of $\max(c_s^2/v_{\rm esc}^2)\simeq0.215\pm0.01$ in the post-shock medium, but with dependencies on $\cal{M}$ and $\dot{M}$. The maximum critical antesonic ratio changes by $\simeq 3-5$\% depending on $\mathcal{M}$ across the range of $\dot{M}$ we explore (Figure \ref{fig:maxAntesonicVsMdot}). Pressurized inflow and time-dependent dynamics modify the behavior of the antesonic ratio in a way that is not captured by semi-analytic and time-steady models that assume pressure-less free-fall (e.g., \citealt{Pejcha2012,Raives2018}).  
  
  \item We explore the heuristic timescale ratio as an explosion condition (see Section \ref{subsec:advheat}). We find that the maximum value of the ratio of the advection time to the heating time in the gain region, i.e., the critical timescale ratio, varies along the critical curve. We also show that the critical timescale ratio depends on the definition of the advection and heating timescales. We find that this ratio is below unity for near-critical stable models, it increases with $\mathcal{M}$, and it converges toward the limit of pressureless free-fall (see Section \ref{subsec:advheat} and Figure \ref{fig:AllMaxTimescalesBoundaryValues}). Depending on $\mathcal{M}$, the critical timescale ratio varies by a factor of two or three along the critical curve.
  
  \item We also explore the integral force explosion condition (FEC), as quantified by $\tilde{\Psi}$ (equation \ref{eqn:ForceExplosionCondition}). As with other explosions conditions, the critical value of $\tilde{\Psi}$ is a function of  both $\mathcal{M}$ and $\dot{M}$ (see Section \ref{subsec:FEC}), but it converges as the limit of pressureless free-fall is approached (See Figures \ref{fig:PsiVst} and \ref{fig:NearCriticalPsi}). Depending on $\mathcal{M}$, the critical value of $\tilde{\Psi}$ varies by about a factor of $\simeq 0.2 - 2$ across the range of $\dot{M}$ we explore.
  
  \item Using our fiducial inputs, the critical antesonic ratio shows the least variation along the critical curve ($\simeq 3-5 \%$ depending on $\mathcal{M}$) compared to the heuristic timescale condition (a factor of 2 to 3 depending on $\mathcal{M}$) and the FEC (a factor of $0.2$ to $2$ depending on $\mathcal{M}$).

  \item We find shock oscillations for stable accretion models at low $\dot{M}$ that complicate the identification of a single explosion condition. Shock oscillations produce corresponding changes in all diagnostic quantities. For example, they may generate maximum antesonic ratios that are temporarily above the critical value (see Figure \ref{fig:maxAntesonicVsMdot}). Similar variations occur in the heuristic timescale condition and in the FEC. 
    
\end{itemize}

With this computational framework in place, our future work will address 2D and 3D models, where we can assess the role of multidimensional instabilities in changing the normalization and character of the critical condition. Further work must also be done to understand the role of $\mathcal{M}$ in mapping realistic progenitor structures to the critical condition.

\section*{Acknowledgments}
\label{section:acknowledgements}
We thank Tejas Prasanna and Matthias Raives for helpful discussions. We thank Matt Colemann for helpful input regarding Athena++. We acknowledge partial support from NASA grant 80NSSC20K0531.

\section*{Data Availability}
The implementation of the EoS, the problem generator file to run the simulations using Athena++, and the data shown in this paper are available upon request.

\bibliographystyle{mnras}
\bibliography{paper_main} 

\begin{thebibliography}{}
\makeatletter
\relax
\def\mn@urlcharsother{\let\do\@makeother \do\$\do\&\do\#\do\^\do\_\do\%\do\~}
\def\mn@doi{\begingroup\mn@urlcharsother \@ifnextchar [ {\mn@doi@} {\mn@doi@[]}}
\def\mn@doi@[#1]#2{\def\@tempa{#1}\ifx\@tempa\@empty \href {http://dx.doi.org/#2} {doi:#2}\else \href {http://dx.doi.org/#2} {#1}\fi \endgroup}
\def\mn@eprint#1#2{\mn@eprint@#1:#2::\@nil}
\def\mn@eprint@arXiv#1{\href {http://arxiv.org/abs/#1} {{\tt arXiv:#1}}}
\def\mn@eprint@dblp#1{\href {http://dblp.uni-trier.de/rec/bibtex/#1.xml} {dblp:#1}}
\def\mn@eprint@#1:#2:#3:#4\@nil{\def\@tempa {#1}\def\@tempb {#2}\def\@tempc {#3}\ifx \@tempc \@empty \let \@tempc \@tempb \let \@tempb \@tempa \fi \ifx \@tempb \@empty \def\@tempb {arXiv}\fi \@ifundefined {mn@eprint@\@tempb}{\@tempb:\@tempc}{\expandafter \expandafter \csname mn@eprint@\@tempb\endcsname \expandafter{\@tempc}}}

\bibitem[\protect\citeauthoryear{Adams, Kochanek, Gerke, Stanek  \& Dai}{Adams et~al.}{2017a}]{adams2017a}
Adams S.,  Kochanek C.,  Gerke J.,  Stanek K.,   Dai X.,  2017a, Monthly Notices of The Royal Astronomical Society, 468, 4968

\bibitem[\protect\citeauthoryear{Adams, Kochanek, Gerke  \& Stanek}{Adams et~al.}{2017b}]{adams2017b}
Adams S.,  Kochanek C.,  Gerke J.,   Stanek K.,  2017b, Monthly Notices of the Royal Astronomical Society, 469, 1445

\bibitem[\protect\citeauthoryear{Bethe}{Bethe}{1990}]{Bethe1990}
Bethe H.~A.,  1990, \mn@doi [Rev. Mod. Phys.] {10.1103/RevModPhys.62.801}, 62, 801

\bibitem[\protect\citeauthoryear{{Bethe} \& {Wilson}}{{Bethe} \& {Wilson}}{1985}]{Bethe1985}
{Bethe} H.~A.,  {Wilson} J.~R.,  1985, \mn@doi [\apj] {10.1086/163343}, \href {https://ui.adsabs.harvard.edu/abs/1985ApJ...295...14B} {295, 14}

\bibitem[\protect\citeauthoryear{{Bruenn}, {Mezzacappa}  \& {Dineva}}{{Bruenn} et~al.}{1995}]{Bruenn1995}
{Bruenn} S.~W.,  {Mezzacappa} A.,   {Dineva} T.,  1995, \mn@doi [\physrep] {10.1016/0370-1573(94)00102-9}, \href {https://ui.adsabs.harvard.edu/abs/1995PhR...256...69B} {256, 69}

\bibitem[\protect\citeauthoryear{{Burrows}}{{Burrows}}{2013}]{Burrows2013}
{Burrows} A.,  2013, \mn@doi [Reviews of Modern Physics] {10.1103/RevModPhys.85.245}, \href {https://ui.adsabs.harvard.edu/abs/2013RvMP...85..245B} {85, 245}

\bibitem[\protect\citeauthoryear{{Burrows} \& {Goshy}}{{Burrows} \& {Goshy}}{1993}]{Burrows1993}
{Burrows} A.,  {Goshy} J.,  1993, \mn@doi [\apjl] {10.1086/187074}, \href {https://ui.adsabs.harvard.edu/abs/1993ApJ...416L..75B} {416, L75}

\bibitem[\protect\citeauthoryear{{Burrows} \& {Lattimer}}{{Burrows} \& {Lattimer}}{1985}]{Burrows1985}
{Burrows} A.,  {Lattimer} J.~M.,  1985, \mn@doi [\apjl] {10.1086/184572}, \href {https://ui.adsabs.harvard.edu/abs/1985ApJ...299L..19B} {299, L19}

\bibitem[\protect\citeauthoryear{{Burrows} \& {Lattimer}}{{Burrows} \& {Lattimer}}{1986}]{Burrows1986}
{Burrows} A.,  {Lattimer} J.~M.,  1986, \mn@doi [\apj] {10.1086/164405}, \href {https://ui.adsabs.harvard.edu/abs/1986ApJ...307..178B} {307, 178}

\bibitem[\protect\citeauthoryear{{Burrows} \& {Vartanyan}}{{Burrows} \& {Vartanyan}}{2021}]{Burrows2021}
{Burrows} A.,  {Vartanyan} D.,  2021, \mn@doi [\nat] {10.1038/s41586-020-03059-w}, \href {https://ui.adsabs.harvard.edu/abs/2021Natur.589...29B} {589, 29}

\bibitem[\protect\citeauthoryear{{Burrows}, {Radice}, {Vartanyan}, {Nagakura}, {Skinner}  \& {Dolence}}{{Burrows} et~al.}{2020}]{Burrows2020a}
{Burrows} A.,  {Radice} D.,  {Vartanyan} D.,  {Nagakura} H.,  {Skinner} M.~A.,   {Dolence} J.~C.,  2020, \mn@doi [\mnras] {10.1093/mnras/stz3223}, \href {https://ui.adsabs.harvard.edu/abs/2020MNRAS.491.2715B} {491, 2715}

\bibitem[\protect\citeauthoryear{{Burrows}, {Wang}  \& {Vartanyan}}{{Burrows} et~al.}{2024}]{Burrows2024}
{Burrows} A.,  {Wang} T.,   {Vartanyan} D.,  2024, \mn@doi [\apjl] {10.3847/2041-8213/ad319e}, \href {https://ui.adsabs.harvard.edu/abs/2024ApJ...964L..16B} {964, L16}

\bibitem[\protect\citeauthoryear{{Colella}}{{Colella}}{1990}]{Collela1990}
{Colella} P.,  1990, \mn@doi [Journal of Computational Physics] {10.1016/0021-9991(90)90233-Q}, \href {https://ui.adsabs.harvard.edu/abs/1990JCoPh..87..171C} {87, 171}

\bibitem[\protect\citeauthoryear{Colella \& Woodward}{Colella \& Woodward}{1984}]{Colella1984}
Colella P.,  Woodward P.~R.,  1984, Journal of computational physics, 54, 174

\bibitem[\protect\citeauthoryear{Coleman}{Coleman}{2020}]{Coleman2020}
Coleman M.~S.,  2020, The Astrophysical Journal Supplement Series, 248, 7

\bibitem[\protect\citeauthoryear{{Colgate}, {Grasberger}  \& {White}}{{Colgate} et~al.}{1961}]{Colgate1961}
{Colgate} S.~A.,  {Grasberger} W.~H.,   {White} R.~H.,  1961, \mn@doi [\aj] {10.1086/108573}, \href {https://ui.adsabs.harvard.edu/abs/1961AJ.....66S.280C} {66, 280}

\bibitem[\protect\citeauthoryear{{Couch}}{{Couch}}{2013}]{Couch2013}
{Couch} S.~M.,  2013, \mn@doi [\apj] {10.1088/0004-637X/775/1/35}, \href {https://ui.adsabs.harvard.edu/abs/2013ApJ...775...35C} {775, 35}

\bibitem[\protect\citeauthoryear{Couch}{Couch}{2017}]{couch2017}
Couch S.~M.,  2017, Philosophical Transactions of the Royal Society A: Mathematical, Physical and Engineering Sciences, 375, 20160271

\bibitem[\protect\citeauthoryear{{Couch} \& {Ott}}{{Couch} \& {Ott}}{2015}]{Couch2015}
{Couch} S.~M.,  {Ott} C.~D.,  2015, \mn@doi [\apj] {10.1088/0004-637X/799/1/5}, \href {https://ui.adsabs.harvard.edu/abs/2015ApJ...799....5C} {799, 5}

\bibitem[\protect\citeauthoryear{De~Moura \& Kubrusly}{De~Moura \& Kubrusly}{2013}]{DeMoura2013}
De~Moura C.~A.,  Kubrusly C.~S.,  2013, AMC, 10

\bibitem[\protect\citeauthoryear{{Fern{\'a}ndez}}{{Fern{\'a}ndez}}{2012}]{Fernandez2012}
{Fern{\'a}ndez} R.,  2012, \mn@doi [\apj] {10.1088/0004-637X/749/2/142}, \href {https://ui.adsabs.harvard.edu/abs/2012ApJ...749..142F} {749, 142}

\bibitem[\protect\citeauthoryear{{Fern{\'a}ndez}}{{Fern{\'a}ndez}}{2015}]{2015Fernandez}
{Fern{\'a}ndez} R.,  2015, \mn@doi [\mnras] {10.1093/mnras/stv146310.48550/arXiv.1504.07996}, \href {https://ui.adsabs.harvard.edu/abs/2015MNRAS.452.2071F} {452, 2071}

\bibitem[\protect\citeauthoryear{Fern{\'a}ndez \& Thompson}{Fern{\'a}ndez \& Thompson}{2009}]{Fernandez2009}
Fern{\'a}ndez R.,  Thompson C.,  2009, The Astrophysical Journal, 697, 1827

\bibitem[\protect\citeauthoryear{Fern{\'a}ndez, M{\"u}ller, Foglizzo  \& Janka}{Fern{\'a}ndez et~al.}{2014}]{Fernandez2014}
Fern{\'a}ndez R.,  M{\"u}ller B.,  Foglizzo T.,   Janka H.-T.,  2014, Monthly Notices of the Royal Astronomical Society, 440, 2763

\bibitem[\protect\citeauthoryear{Ferreira \& Provid\^encia}{Ferreira \& Provid\^encia}{2021}]{Ferreira2021}
Ferreira M.,  Provid\^encia C. m.~c.,  2021, \mn@doi [Phys. Rev. D] {10.1103/PhysRevD.104.063006}, 104, 063006

\bibitem[\protect\citeauthoryear{{Gabay}, {Balberg}  \& {Keshet}}{{Gabay} et~al.}{2015}]{Gabey2015}
{Gabay} D.,  {Balberg} S.,   {Keshet} U.,  2015, \mn@doi [\apj] {10.1088/0004-637X/815/1/37}, \href {https://ui.adsabs.harvard.edu/abs/2015ApJ...815...37G} {815, 37}

\bibitem[\protect\citeauthoryear{Gerke, Kochanek  \& Stanek}{Gerke et~al.}{2015}]{gerke2015}
Gerke J.,  Kochanek C.,   Stanek K.,  2015, Monthly Notices of the Royal Astronomical Society, 450, 3289

\bibitem[\protect\citeauthoryear{{Gogilashvili} \& {Murphy}}{{Gogilashvili} \& {Murphy}}{2022}]{Gogilashvili2022}
{Gogilashvili} M.,  {Murphy} J.~W.,  2022, \mn@doi [\mnras] {10.1093/mnras/stac1811}, \href {https://ui.adsabs.harvard.edu/abs/2022MNRAS.515.1610G} {515, 1610}

\bibitem[\protect\citeauthoryear{{Gogilashvili}, {Murphy}  \& {O'Connor}}{{Gogilashvili} et~al.}{2023}]{Gogilashvili2023}
{Gogilashvili} M.,  {Murphy} J.~W.,   {O'Connor} E.~P.,  2023, \mn@doi [\mnras] {10.1093/mnras/stad2155}, \href {https://ui.adsabs.harvard.edu/abs/2023MNRAS.524.4109G} {524, 4109}

\bibitem[\protect\citeauthoryear{{Hanke}, {Marek}, {M{\"u}ller}  \& {Janka}}{{Hanke} et~al.}{2012}]{Hanke2012}
{Hanke} F.,  {Marek} A.,  {M{\"u}ller} B.,   {Janka} H.-T.,  2012, \mn@doi [\apj] {10.1088/0004-637X/755/2/138}, \href {https://ui.adsabs.harvard.edu/abs/2012ApJ...755..138H} {755, 138}

\bibitem[\protect\citeauthoryear{Hanke, M{\"u}ller, Wongwathanarat, Marek  \& Janka}{Hanke et~al.}{2013}]{hanke2013}
Hanke F.,  M{\"u}ller B.,  Wongwathanarat A.,  Marek A.,   Janka H.-T.,  2013, The Astrophysical Journal, 770, 66

\bibitem[\protect\citeauthoryear{{Janka}}{{Janka}}{2000}]{Janka2000}
{Janka} H.~T.,  2000, \mn@doi [\nphysa] {10.1016/S0375-9474(99)00580-1}, \href {https://ui.adsabs.harvard.edu/abs/2000NuPhA.663..119J} {663, 119}

\bibitem[\protect\citeauthoryear{{Janka}}{{Janka}}{2012}]{Janka2012}
{Janka} H.-T.,  2012, \mn@doi [Annual Review of Nuclear and Particle Science] {10.1146/annurev-nucl-102711-094901}, \href {https://ui.adsabs.harvard.edu/abs/2012ARNPS..62..407J} {62, 407}

\bibitem[\protect\citeauthoryear{{Janka}, {Melson}  \& {Summa}}{{Janka} et~al.}{2016}]{Janka2016}
{Janka} H.-T.,  {Melson} T.,   {Summa} A.,  2016, \mn@doi [Annual Review of Nuclear and Particle Science] {10.1146/annurev-nucl-102115-044747}, \href {https://ui.adsabs.harvard.edu/abs/2016ARNPS..66..341J} {66, 341}

\bibitem[\protect\citeauthoryear{Kochanek, Beacom, Kistler, Prieto, Stanek, Thompson  \& Y{\"u}ksel}{Kochanek et~al.}{2008}]{kochanek2008}
Kochanek C.~S.,  Beacom J.~F.,  Kistler M.~D.,  Prieto J.~L.,  Stanek K.~Z.,  Thompson T.~A.,   Y{\"u}ksel H.,  2008, The Astrophysical Journal, 684, 1336

\bibitem[\protect\citeauthoryear{{Kotake}, {Ohnishi}  \& {Yamada}}{{Kotake} et~al.}{2007}]{Kotake2007}
{Kotake} K.,  {Ohnishi} N.,   {Yamada} S.,  2007, \mn@doi [\apj] {10.1086/509320}, \href {https://ui.adsabs.harvard.edu/abs/2007ApJ...655..406K} {655, 406}

\bibitem[\protect\citeauthoryear{{Lentz} et~al.,}{{Lentz} et~al.}{2015}]{Lentz2015}
{Lentz} E.~J.,  et~al., 2015, \mn@doi [\apjl] {10.1088/2041-8205/807/2/L31}, \href {https://ui.adsabs.harvard.edu/abs/2015ApJ...807L..31L} {807, L31}

\bibitem[\protect\citeauthoryear{{Mabanta} \& {Murphy}}{{Mabanta} \& {Murphy}}{2018}]{Mabanta2018}
{Mabanta} Q.~A.,  {Murphy} J.~W.,  2018, \mn@doi [\apj] {10.3847/1538-4357/aaaec7}, \href {https://ui.adsabs.harvard.edu/abs/2018ApJ...856...22M} {856, 22}

\bibitem[\protect\citeauthoryear{{Mazurek}}{{Mazurek}}{1982}]{Mazurek1982}
{Mazurek} T.~J.,  1982, \mn@doi [\apjl] {10.1086/183839}, \href {https://ui.adsabs.harvard.edu/abs/1982ApJ...259L..13M} {259, L13}

\bibitem[\protect\citeauthoryear{{Melson}, {Janka}, {Bollig}, {Hanke}, {Marek}  \& {M{\"u}ller}}{{Melson} et~al.}{2015}]{Melson2015}
{Melson} T.,  {Janka} H.-T.,  {Bollig} R.,  {Hanke} F.,  {Marek} A.,   {M{\"u}ller} B.,  2015, \mn@doi [\apjl] {10.1088/2041-8205/808/2/L42}, \href {https://ui.adsabs.harvard.edu/abs/2015ApJ...808L..42M} {808, L42}

\bibitem[\protect\citeauthoryear{{Mezzacappa}, {Endeve}, {Messer}  \& {Bruenn}}{{Mezzacappa} et~al.}{2020}]{Mezz2020}
{Mezzacappa} A.,  {Endeve} E.,  {Messer} O.~E.~B.,   {Bruenn} S.~W.,  2020, \mn@doi [Living Reviews in Computational Astrophysics] {10.1007/s41115-020-00010-8}, \href {https://ui.adsabs.harvard.edu/abs/2020LRCA....6....4M} {6, 4}

\bibitem[\protect\citeauthoryear{{M{\"u}ller}, {Melson}, {Heger}  \& {Janka}}{{M{\"u}ller} et~al.}{2017}]{Muller2017}
{M{\"u}ller} B.,  {Melson} T.,  {Heger} A.,   {Janka} H.-T.,  2017, \mn@doi [\mnras] {10.1093/mnras/stx1962}, \href {https://ui.adsabs.harvard.edu/abs/2017MNRAS.472..491M} {472, 491}

\bibitem[\protect\citeauthoryear{{Murphy} \& {Burrows}}{{Murphy} \& {Burrows}}{2008}]{Murphy2008}
{Murphy} J.~W.,  {Burrows} A.,  2008, \mn@doi [\apj] {10.1086/592214}, \href {https://ui.adsabs.harvard.edu/abs/2008ApJ...688.1159M} {688, 1159}

\bibitem[\protect\citeauthoryear{Murphy \& Dolence}{Murphy \& Dolence}{2017}]{Murphy2017}
Murphy J.~W.,  Dolence J.~C.,  2017, The Astrophysical Journal, 834, 183

\bibitem[\protect\citeauthoryear{Neustadt, Kochanek, Stanek, Basinger, Jayasinghe, Garling, Adams  \& Gerke}{Neustadt et~al.}{2021}]{neustadt2021}
Neustadt J.,  Kochanek C.,  Stanek K.,  Basinger C.,  Jayasinghe T.,  Garling C.,  Adams S.,   Gerke J.,  2021, Monthly Notices of the Royal Astronomical Society, 508, 516

\bibitem[\protect\citeauthoryear{Nordhaus, Brandt, Burrows  \& Almgren}{Nordhaus et~al.}{2012}]{Nordhaus2010}
Nordhaus J.,  Brandt T.~D.,  Burrows A.,   Almgren A.,  2012, \mn@doi [Monthly Notices of the Royal Astronomical Society] {10.1111/j.1365-2966.2012.21002.x}, 423, 1805

\bibitem[\protect\citeauthoryear{{Ott}, {O'Connor}, {Gossan}, {Abdikamalov}, {Gamma}  \& {Drasco}}{{Ott} et~al.}{2013}]{Ott2013}
{Ott} C.~D.,  {O'Connor} E.~P.,  {Gossan} S.,  {Abdikamalov} E.,  {Gamma} U.~C.~T.,   {Drasco} S.,  2013, \mn@doi [Nuclear Physics B Proceedings Supplements] {10.1016/j.nuclphysbps.2013.04.036}, \href {https://ui.adsabs.harvard.edu/abs/2013NuPhS.235..381O} {235, 381}

\bibitem[\protect\citeauthoryear{Ott, Roberts, da Silva~Schneider, Fedrow, Haas  \& Schnetter}{Ott et~al.}{2018b}]{Ott2018a}
Ott C.~D.,  Roberts L.~F.,  da Silva~Schneider A.,  Fedrow J.~M.,  Haas R.,   Schnetter E.,  2018b, \mn@doi [The Astrophysical Journal] {10.3847/2041-8213/aaa967}, 855, L3

\bibitem[\protect\citeauthoryear{{Ott}, {Roberts}, {da Silva Schneider}, {Fedrow}, {Haas}  \& {Schnetter}}{{Ott} et~al.}{2018a}]{Ott2018b}
{Ott} C.~D.,  {Roberts} L.~F.,  {da Silva Schneider} A.,  {Fedrow} J.~M.,  {Haas} R.,   {Schnetter} E.,  2018a, \mn@doi [\apjl] {10.3847/2041-8213/aaa967}, \href {https://ui.adsabs.harvard.edu/abs/2018ApJ...855L...3O} {855, L3}

\bibitem[\protect\citeauthoryear{{Pejcha} \& {Thompson}}{{Pejcha} \& {Thompson}}{2012}]{Pejcha2012}
{Pejcha} O.,  {Thompson} T.~A.,  2012, \mn@doi [\apj] {10.1088/0004-637X/746/1/106}, \href {https://ui.adsabs.harvard.edu/abs/2012ApJ...746..106P} {746, 106}

\bibitem[\protect\citeauthoryear{{Prasanna}, {Coleman}, {Raives}  \& {Thompson}}{{Prasanna} et~al.}{2022}]{Prasanna2022}
{Prasanna} T.,  {Coleman} M. S.~B.,  {Raives} M.~J.,   {Thompson} T.~A.,  2022, \mn@doi [\mnras] {10.1093/mnras/stac2651}, \href {https://ui.adsabs.harvard.edu/abs/2022MNRAS.517.3008P} {517, 3008}

\bibitem[\protect\citeauthoryear{{Qian} \& {Woosley}}{{Qian} \& {Woosley}}{1996}]{Qian1996}
{Qian} Y.~Z.,  {Woosley} S.~E.,  1996, \mn@doi [\apj] {10.1086/177973}, \href {https://ui.adsabs.harvard.edu/abs/1996ApJ...471..331Q} {471, 331}

\bibitem[\protect\citeauthoryear{{Raives}, {Couch}, {Greco}, {Pejcha}  \& {Thompson}}{{Raives} et~al.}{2018}]{Raives2018}
{Raives} M.~J.,  {Couch} S.~M.,  {Greco} J.~P.,  {Pejcha} O.,   {Thompson} T.~A.,  2018, \mn@doi [\mnras] {10.1093/mnras/sty2457}, \href {https://ui.adsabs.harvard.edu/abs/2018MNRAS.481.3293R} {481, 3293}

\bibitem[\protect\citeauthoryear{{Raives}, {Thompson}  \& {Couch}}{{Raives} et~al.}{2021}]{Raives2021}
{Raives} M.~J.,  {Thompson} T.~A.,   {Couch} S.~M.,  2021, \mn@doi [\mnras] {10.1093/mnras/stab286}, \href {https://ui.adsabs.harvard.edu/abs/2021MNRAS.502.4125R} {502, 4125}

\bibitem[\protect\citeauthoryear{{Scheck}, {Kifonidis}, {Janka}  \& {M{\"u}ller}}{{Scheck} et~al.}{2006}]{Scheck2006}
{Scheck} L.,  {Kifonidis} K.,  {Janka} H.~T.,   {M{\"u}ller} E.,  2006, \mn@doi [\aap] {10.1051/0004-6361:20064855}, \href {https://ui.adsabs.harvard.edu/abs/2006A&A...457..963S} {457, 963}

\bibitem[\protect\citeauthoryear{{Stockinger} et~al.,}{{Stockinger} et~al.}{2020}]{Stockinger2020}
{Stockinger} G.,  et~al., 2020, \mn@doi [\mnras] {10.1093/mnras/staa1691}, \href {https://ui.adsabs.harvard.edu/abs/2020MNRAS.496.2039S} {496, 2039}

\bibitem[\protect\citeauthoryear{Stone \& Gardiner}{Stone \& Gardiner}{2009}]{Stone2009}
Stone J.~M.,  Gardiner T.,  2009, New Astronomy, 14, 139

\bibitem[\protect\citeauthoryear{{Stone}, {Tomida}, {White}  \& {Felker}}{{Stone} et~al.}{2020}]{Stone2020}
{Stone} J.~M.,  {Tomida} K.,  {White} C.~J.,   {Felker} K.~G.,  2020, \mn@doi [\apjs] {10.3847/1538-4365/ab929b}, \href {https://ui.adsabs.harvard.edu/abs/2020ApJS..249....4S} {249, 4}

\bibitem[\protect\citeauthoryear{Sukhbold, Ertl, Woosley, Brown  \& Janka}{Sukhbold et~al.}{2016}]{sukhbold2016}
Sukhbold T.,  Ertl T.,  Woosley S.,  Brown J.~M.,   Janka H.-T.,  2016, The Astrophysical Journal, 821, 38

\bibitem[\protect\citeauthoryear{Sukhbold, Woosley  \& Heger}{Sukhbold et~al.}{2018}]{sukhbold2018}
Sukhbold T.,  Woosley S.,   Heger A.,  2018, The Astrophysical Journal, 860, 93

\bibitem[\protect\citeauthoryear{{Summa}, {Janka}, {Melson}  \& {Marek}}{{Summa} et~al.}{2018}]{Summa2018}
{Summa} A.,  {Janka} H.-T.,  {Melson} T.,   {Marek} A.,  2018, \mn@doi [\apj] {10.3847/1538-4357/aa9ce8}, \href {https://ui.adsabs.harvard.edu/abs/2018ApJ...852...28S} {852, 28}

\bibitem[\protect\citeauthoryear{Suwa, Yamada, Takiwaki  \& Kotake}{Suwa et~al.}{2015}]{suwa2015}
Suwa Y.,  Yamada S.,  Takiwaki T.,   Kotake K.,  2015, The Astrophysical Journal, 816, 43

\bibitem[\protect\citeauthoryear{{Takiwaki}, {Kotake}  \& {Suwa}}{{Takiwaki} et~al.}{2014}]{Takiwaki2014}
{Takiwaki} T.,  {Kotake} K.,   {Suwa} Y.,  2014, \mn@doi [\apj] {10.1088/0004-637X/786/2/83}, \href {https://ui.adsabs.harvard.edu/abs/2014ApJ...786...83T} {786, 83}

\bibitem[\protect\citeauthoryear{{Thompson}}{{Thompson}}{2000}]{ThompsonC2000}
{Thompson} C.,  2000, \mn@doi [\apj] {10.1086/308773}, \href {https://ui.adsabs.harvard.edu/abs/2000ApJ...534..915T} {534, 915}

\bibitem[\protect\citeauthoryear{Thompson, Burrows  \& Meyer}{Thompson et~al.}{2001}]{Thompson2001}
Thompson T.~A.,  Burrows A.,   Meyer B.~S.,  2001, The Astrophysical Journal, 562, 887

\bibitem[\protect\citeauthoryear{Thompson, Burrows  \& Pinto}{Thompson et~al.}{2003}]{thompson2003}
Thompson T.~A.,  Burrows A.,   Pinto P.~A.,  2003, The Astrophysical Journal, 592, 434

\bibitem[\protect\citeauthoryear{Thompson, Quataert  \& Burrows}{Thompson et~al.}{2005}]{Thompson2005}
Thompson T.~A.,  Quataert E.,   Burrows A.,  2005, The Astrophysical Journal, 620, 861

\bibitem[\protect\citeauthoryear{Timmes \& Swesty}{Timmes \& Swesty}{2000}]{Timmes2000}
Timmes F.~X.,  Swesty F.~D.,  2000, The Astrophysical Journal Supplement Series, 126, 501

\bibitem[\protect\citeauthoryear{Toro, Spruce  \& Speares}{Toro et~al.}{1994}]{Toro1994}
Toro E.~F.,  Spruce M.,   Speares W.,  1994, Shock waves, 4, 25

\bibitem[\protect\citeauthoryear{Vartanyan, Burrows, Radice, Skinner  \& Dolence}{Vartanyan et~al.}{2018}]{vartanyan2018}
Vartanyan D.,  Burrows A.,  Radice D.,  Skinner M.~A.,   Dolence J.,  2018, Monthly Notices of the Royal Astronomical Society, 477, 3091

\bibitem[\protect\citeauthoryear{{Wang} \& {Burrows}}{{Wang} \& {Burrows}}{2024}]{WangBurrows2024}
{Wang} T.,  {Burrows} A.,  2024, \mn@doi [\apj] {10.3847/1538-4357/ad12b8}, \href {https://ui.adsabs.harvard.edu/abs/2024ApJ...962...71W} {962, 71}

\bibitem[\protect\citeauthoryear{Wang, Vartanyan, Burrows  \& Coleman}{Wang et~al.}{2022}]{Wang2022}
Wang T.,  Vartanyan D.,  Burrows A.,   Coleman M.~S.,  2022, Monthly Notices of the Royal Astronomical Society, 517, 543

\bibitem[\protect\citeauthoryear{{Woosley} \& {Heger}}{{Woosley} \& {Heger}}{2007}]{Woosley2007}
{Woosley} S.~E.,  {Heger} A.,  2007, \mn@doi [\physrep] {10.1016/j.physrep.2007.02.009}, \href {https://ui.adsabs.harvard.edu/abs/2007PhR...442..269W} {442, 269}

\bibitem[\protect\citeauthoryear{Woosley, Heger  \& Weaver}{Woosley et~al.}{2002}]{Woosley2002}
Woosley S.~E.,  Heger A.,   Weaver T.~A.,  2002, \mn@doi [Rev. Mod. Phys.] {10.1103/RevModPhys.74.1015}, 74, 1015

\bibitem[\protect\citeauthoryear{{Yamasaki} \& {Yamada}}{{Yamasaki} \& {Yamada}}{2005}]{Yamasaki2005}
{Yamasaki} T.,  {Yamada} S.,  2005, \mn@doi [\apj] {10.1086/428496}, \href {https://ui.adsabs.harvard.edu/abs/2005ApJ...623.1000Y} {623, 1000}

\bibitem[\protect\citeauthoryear{{Yamasaki} \& {Yamada}}{{Yamasaki} \& {Yamada}}{2006}]{Yamasaki2006}
{Yamasaki} T.,  {Yamada} S.,  2006, \mn@doi [\apj] {10.1086/507067}, \href {https://ui.adsabs.harvard.edu/abs/2006ApJ...650..291Y} {650, 291}

\makeatother
\end{thebibliography}


\label{lastpage}
\end{document}